\documentclass[letterpaper,english,american,prr,twocolumn,amsmath,amssymb,superscriptaddress,nofootinbib]{revtex4-2}
\usepackage[T1]{fontenc}
\usepackage{babel}
\usepackage{prettyref}
\usepackage{float}
\usepackage{units}
\usepackage{mathtools}
\usepackage{amsmath}
\usepackage{amssymb}
\usepackage{graphicx}
\usepackage[unicode=true,
 bookmarks=true,bookmarksnumbered=false,bookmarksopen=false,
 breaklinks=false,pdfborder={0 0 1},backref=false,colorlinks=false]
 {hyperref}
\hypersetup{pdftitle={Field theory for optimal signal propagation in ResNets},
 pdfauthor={Kirsten Fischer, David Dahmen, Moritz Helias},
 colorlinks=true,linkcolor=black,citecolor=black,urlcolor=black,filecolor=black}

\makeatletter

\pdfpageheight\paperheight
\pdfpagewidth\paperwidth

\usepackage[latin1]{inputenc}

\newrefformat{eq}{\hyperref[#1]{Eq.~(\ref{#1})}}
\newrefformat{cap}{\hyperref[#1]{Fig.~\ref{#1}}}
\newrefformat{fig}{\hyperref[#1]{Fig.~\ref{#1}}}
\newrefformat{tab}{\hyperref[#1]{Table ~\ref{#1}}}
\newrefformat{sec}{\hyperref[#1]{Section~\ref{#1}}}
\newrefformat{subsec}{\hyperref[#1]{Section~\ref{#1}}}
\newrefformat{cha}{\hyperref[#1]{Chapter~\ref{#1}}}
\newrefformat{app}{\hyperref[#1]{Appendix~\ref{#1}}}

\usepackage{MnSymbol}
\DeclareMathAlphabet\mathbb{U}{msb}{m}{n}

\usepackage{booktabs}
\usepackage[title]{appendix}
\usepackage{aligned-overset}

\usepackage[force]{feynmp-auto}

\makeatother

\begin{document}
\selectlanguage{english}%
 \renewcommand{\figurename}{FIG.} \renewcommand{\tablename}{TABLE} \renewcommand{\appendixname}{APPENDIX}
\global\long\def\R{\mathbb{R}}%
\global\long\def\llangle{\langle\!\langle}%
\global\long\def\rrangle{\rangle\!\rangle}%
\global\long\def\T{\mathsf{T}}%

\title{\selectlanguage{english}%
Field theory for optimal signal propagation in ResNets}
\author{\selectlanguage{english}%
Kirsten Fischer}
\email{ki.fischer@fz-juelich.de}

\affiliation{\selectlanguage{english}%
Institute for Advanced Simulation (IAS-6), Jülich Research Centre,
Jülich, Germany}
\affiliation{\selectlanguage{english}%
RWTH Aachen University, Aachen, Germany}
\author{\selectlanguage{english}%
David Dahmen}
\affiliation{\selectlanguage{english}%
Institute for Advanced Simulation (IAS-6), Jülich Research Centre,
Jülich, Germany}
\author{\selectlanguage{english}%
Moritz Helias}
\affiliation{\selectlanguage{english}%
Institute for Advanced Simulation (IAS-6), Jülich Research Centre,
Jülich, Germany}
\affiliation{\selectlanguage{english}%
Department of Physics, Faculty 1, RWTH Aachen University, Aachen,
Germany}
\date{\selectlanguage{english}%
\today}
\begin{abstract}
Residual networks have significantly better trainability and thus
performance than feed-forward networks at large depth. Introducing
skip connections facilitates signal propagation to deeper layers.
In addition, previous works found that adding a scaling parameter
for the residual branch further improves generalization performance.
While they empirically identified a particularly beneficial range
of values for this scaling parameter, the mechanism for the resulting
performance improvement and its universality across network hyperparameters
remain an open question. For feed-forward networks, finite-size theories
have led to important insights with regard to signal propagation and
hyperparameter tuning. We here derive a systematic finite-size field
theory for residual networks to study signal propagation and its
dependence on the scaling for the residual branch. We derive analytical
expressions for the response function, a measure for the network's
sensitivity to inputs, and show that for deep networks the empirically
found values for the scaling parameter lie within the range of maximal
sensitivity. Furthermore, we obtain an analytical expression for the
optimal scaling parameter that depends only weakly on other network
hyperparameters, such as the weight variance, thereby explaining
its universality across hyperparameters. Overall, this work provides
a theoretical framework to study ResNets at finite size.
\end{abstract}
\maketitle

\section{Introduction\label{sec:introduction}}

While feed-forward neural networks (FFNets) have proven successful
at learning a multitude of tasks \citep{Krizhevsky12_1097,Silver16_484},
they become difficult to train at great depths \citep{He16_CVPR}.
As a result, very deep FFNets yield worse performance than their shallow
counterparts. This empirical result is counterintuitive: Since feed-forward
layers in principle can implement identity mappings, which can be
added to a shallow network to increase its depth while yielding the
same performance, such a performance degradation should not be present.
This finding implies that identity mappings are difficult to learn;
in consequence, \citep{He16_CVPR,He16_630} introduced residual networks
(ResNets) that contain skip connections between adjacent layers which
implement identity mappings. Networks such as ResNet-50 \citep{He16_CVPR}
or ResNet-1001 \citep{He16_630} yield state-of-the-art performance
on common benchmark data sets such as CIFAR-10 \citep{Krizhevsky09}.

A scaling of the residual branch, i.e. of the non-identity mapping
in each residual layer, was first introduced by \citet{Szegedy17_4278}
who found that for networks with large numbers of convolutional filters
training becomes unstable and leads to inactive neurons. While this
effect could not be mitigated by additional batch normalization \citep{Ioffe15_448},
downscaling the residual branch by a value $\rho$ between $0.1$
and $0.3$ proved to be a reliable solution. Finding the optimal residual
scaling and a mechanistic explanation for its effectiveness remains
an open question.

One line of research studies how such an optimal residual scaling
depends on the network depth. In the limit of infinite width, ResNets
behave as a Gaussian process. The covariance of this process is also
termed ``Neural Network Gaussian Process (NNGP) kernel'' and has
been derived in \citep{Huang20_33}. While the NNGP corresponds to
training only the readout layer, training all network layers in the
limit of vanishing learning rates leads to a different Gaussian process,
the ``Neural Tangent Kernel'' (NTK) \citep{Jacot18_8580}. In this
kernel perspective, \citet{Huang20_33} argue that the NTK in the
double limit of infinite width and depth becomes degenerate for FFNets
but not for ResNets, suggesting a polynomial scaling of the residual
branch with the inverse depth for better kernel stability at great
depth. \citet{Bachlechner21_1352} include the residual scaling as
a trainable parameter instead of it being a hyperparameter and find
that networks with the residual scaling being initialized at zero
learn an inverse depth scaling.

According to \citet{Tirer22_921}, smaller residual scalings lead
to a smoother NTK and thereby to better interpolation properties between
data points. Studying the spectral properties of the NTK, \citet{Barzilai23}
find a bias of convolutional ResNets towards learning functions with
low-frequency or localized over few pixels. Further, they show that
the scaling proposed by \citet{Huang20_33} leads to a less expressive
dot-product kernel for convolutional ResNets, therefore arguing for
a depth-independent constant residual scaling. By performing a grid
search, \citet{Zhang19_4285} find a value near $0.1$ to yield best
generalization performance for deep ResNets.

\citep{Arpit19_neurips,Hayou21_1324,Hayou21_iclr,Zhang22_3359} argue
for a residual scaling by the square root of the inverse depth: \citet{Arpit19_neurips}
use a mean field analysis to argue that this scaling avoids exploding
or vanishing gradients in the forward and backward pass. While \citep{Hayou21_1324,Hayou21_iclr}
show that the resulting NTK is universal and can express any function,
\citet{Zhang22_3359} find that this scaling stabilizes forward and
backward propagation of the signal in terms of its norm. On the practical
side, \citet{Bordelon24_iclr} show that scaling the residual branch
by the square root of the inverse depth allows them to extend $\mu P$-scaling
\citep{Yang21_iclr}, a particular form of initialization that promotes
trainability, to residual architectures.

We here tackle the problem of optimal scaling from a signal propagation
perspective. We derive a field-theoretic description of residual networks
to study their response function. This function describes the networks'
sensitivity to varying inputs. As the network needs to be able to
distinguish between different data samples, the overall range of output
responses is a relevant indicator for both trainability and generalization.
While a stronger signal generally ensures that two data samples can
be better distinguished, this effect may be counteracted by saturation
effects of the non-linearity in the residual branch of the network.
The residual scaling parameter determines how strongly differences
across data samples are amplified and propagated through the network.

Our main contributions are as follows
\begin{itemize}
\item we derive a novel field-theoretic description of the Bayesian network
prior for residual networks that allows one to systematically account
for finite-size properties of networks;
\item we obtain the response function of residual networks as a finite-size
effect that describes the networks' sensitivity to varying inputs;
\item we show that the response function of the network output as a function
of the residual scaling parameter has a distinct maximum and that
the corresponding optimal residual scaling lies precisely within the
value range empirically found by \citet{Szegedy17_4278};
\item we derive the dependence of the optimal residual scaling on network
hyperparameters and find a strong dependence on the network depth
and weak dependence on all other hyperparameters, explaining the universality
of a $1/\sqrt{\text{depth}}$ scaling for deep residual networks.
\end{itemize}
The field-theoretic framework for the Bayesian network prior can be
generalized beyond ResNets and generally used to systematically take
into account finite-size properties of neural networks. Field-theoretic
descriptions of feed-forward and recurrent networks have been derived
in \citep{Halverson21_035002,Naveh21_064301,Segadlo22_103401,Roberts22}.
Building on a field-theoretic formulation of the Bayesian network
prior, \citep{Naveh21_NeurIPS,ZavatoneVeth21_NeurIPS_I,ZavatoneVeth22_064118,seroussi23_908,Yang23_39380,Fischer24_13660,Rubin24_iclr}
study feature learning in finite-size networks and \citet{Lindner23_arxiv}
study the effect of data variability. \citet{Bordelon23_114009} derive
a dynamical field theory to study gradient dynamics in feed-forward
networks in different scaling regimes. The presented theoretical framework
thus has applications beyond understanding residual scaling, the main
goal of the current work.

The main part is structured into two parts: In \prettyref{sec:theory_resnets},
we first derive a field-theoretic formulation of residual networks.
In this field-theoretic framework, we recover the NNGP as a saddle
point at infinite width and obtain the response function of residual
networks as a finite-width correction to this saddle point. In \prettyref{sec:signal_prop_scaling},
we then study the behavior of the response function as a function
of the residual scaling and find a unique maximum of the response
close to the empirically found optimal values. Finally, we relate
this scaling to optimal signal propagation that is bounded by saturation
effects of the non-linearity and study its dependence on hyperparameters
of the network. The code for theory and experiments can be found in
\citep{Fischer23_zenodo}.

\section{Field Theory of Residual Networks\label{sec:theory_resnets}}

\begin{figure*}
\includegraphics{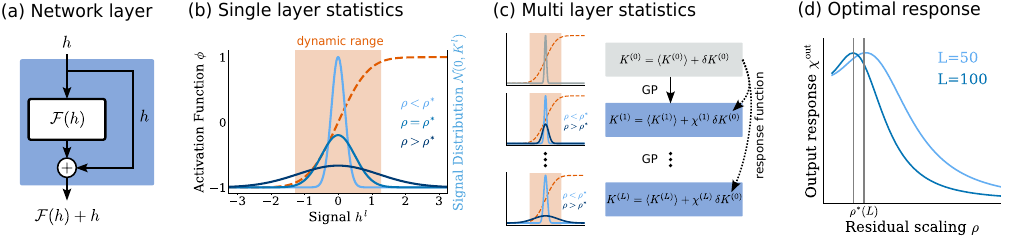}\caption{Signal distribution in residual network. (a) Network layer with residual
branch and skip connection. The residual branch returns $h\protect\mapsto\mathcal{F}(h)$,
the layer passes on $\mathcal{F}(h)+h$ to the next layer. (b) Distribution
of the signal $h^{(l)}$ after layer $l$ (solid curves) relative
to the dynamic range $\mathcal{V}$ (shaded orange area) of the activation
function $\phi=\text{erf}$ (dashed curve). The signal is Gaussian
distributed $h^{(l)}\sim\mathcal{N}(0,K^{(l)})$ with variance given
by $K^{(l)}$, which depends on the residual scaling parameter $\rho$.
For values larger than the optimal scaling $\rho>\rho^{\ast}$, part
of the signal is lost in the saturation of the activation function
$\phi$ (dark blue). For values smaller than the optimal scaling $\rho<\rho^{\ast}$,
the signal is restricted to a small fraction of the dynamic range
(light blue) in which the activation function is typically linear.
For optimal scaling $\rho=\rho^{\ast}$, the signal optimally utilizes
the whole dynamic range $\mathcal{V}$ of the activation function
$\phi$ (blue). (c) The response function $\chi^{(l)}$ describes
how the variance $K^{(l)}$, corresponding to the diagonal element
of the GP kernel, changes to linear order in the perturbation of the
input kernel $\delta K^{(0)}$ around its data mean $\langle K^{(0)}\rangle$.
The kernel $K^{(l)}$ of the signal distribution can only increase
across layers due to the skip connections; its rate of increase is
governed by the residual scaling parameter $\rho$. If the signal
goes into saturation ($\rho>\rho^{\ast}$) or remains close to zero
($\rho<\rho^{\ast}$), then the overall response of the network output
to a change of the input kernel is limited. (d) The output response
$\chi^{\text{out}}$ as a function of the residual scaling $\rho$
exhibits a unique maximum that depends on the network depth $L$,
yielding a scaling $\rho^{\ast}(L)$ that promotes optimal signal
propagation in the network.\label{fig:graphical_abstract}}
\end{figure*}

We here study the following residual architecture
\begin{align}
h^{(0)} & =W^{\text{in}}x+b^{\text{in}},\nonumber \\
h^{(l)} & =h^{(l-1)}+\rho\left[W^{(l)}\phi(h^{(l-1)})+b^{(l)}\right]\quad l=1,\dots,L,\label{eq:network_resnets}\\
y & =W^{\text{out}}\phi(h^{(L)})+b^{\text{out}},\nonumber 
\end{align}
yielding a mapping from the input $x\in\mathbb{R}^{d_{\text{in}}}$
to the output $y\in\mathbb{R}^{d_{\text{out}}}$ as $x\mapsto f(x;\theta)=y$
with trainable network parameters $\theta=\left\{ W^{\text{in}},\,b^{\text{in}},\,W^{(l)},\,b^{(l)},\,W^{\text{out}},\,b^{\text{out}}\right\} $.
Similar to state-of-the-art models such as ResNet-50 \citep{He16_CVPR},
the model contains a linear readin and a fully-connected readout layer.
Thereby, the input $x\in\mathbb{R}^{d_{\text{in}}}$, the signal $h^{(l)}\in\mathbb{R}^{N}$,
and the output $y\in\mathbb{R}^{d_{\text{out}}}$ can have different
dimensions. We refer to the residual branch 
\begin{equation}
\mathcal{F}(h^{(l-1)})\coloneqq\rho\left[W^{(l)}\phi(h^{(l-1)})+b^{(l)}\right]
\end{equation}
together with the skip connection $h^{(l-1)}$ in \eqref{eq:network_resnets}
as a network layer with index $l$ (see \prettyref{fig:graphical_abstract}(a)).
The total number of layers is given by $L$. We assume the non-linear
activation function $\phi$ to be saturating and twice differentiable
almost everywhere; two common choices satisfying both conditions are
the logistic function and the error function. In the following, we
use $\phi=\text{erf}$. The residual branch is multiplied by a scaling
factor $\rho$, which is referred to as the residual scaling parameter
in the following. We study networks at initialization, which is equivalent
to determining the network prior in a setting of Bayesian inference.
To this end, we assume that the network parameters are Gaussian distributed:
For the input layer ${W_{ij}^{\text{in}}\overset{\text{i.i.d.}}{\sim}\mathcal{N}(0,\sigma_{w,\,\text{in}}^{2}/d_{\text{in}}),}\;{b_{i}^{\text{in}}\overset{\text{i.i.d.}}{\sim}\mathcal{N}(0,\sigma_{b,\,\text{in}}^{2}),}\;$for
residual layers $W_{ij}^{(l)}\overset{\text{i.i.d.}}{\sim}\mathcal{N}(0,\sigma_{w}^{2}/N)$,
${b_{i}^{(l)}\overset{\text{i.i.d.}}{\sim}\mathcal{N}(0,\sigma_{b}^{2})},$
and for the readout layer ${W_{ij}^{\text{out}}\overset{\text{i.i.d.}}{\sim}\mathcal{N}(0,\sigma_{w,\,\text{out}}^{2}/N)}$,
${b_{i}^{\text{out}}\overset{\text{i.i.d.}}{\sim}\mathcal{N}(0,\sigma_{b,\,\text{out}}^{2})}$.
The symbol $\sim$ is here and in the following employed as ``is
distributed as''.

\subsection{Network prior in field-theoretic framework\label{subsec:network_prior}}

We derive analytic expressions of the network prior for the residual
network architecture defined in \eqref{eq:network_resnets} in terms
of the network kernels. This derivation uses the field-theoretic approach
employed in \citet{Segadlo22_103401} to study deep feed-forward and
recurrent networks and extends it to residual networks. Given a set
of inputs $X=(x_{\alpha})_{\alpha=1,\dots,P}$, the network prior
defined by
\begin{equation}
p(Y\vert X)\coloneqq\int\mathrm{d}\theta\,p(Y\vert X,\theta)\,p(\theta)
\end{equation}
describes the joint distribution of all outputs ${Y=(y_{\alpha})_{\alpha=1,\dots,P}}$
averaged over all possible initializations of the network parameters
$\theta$; each output $y_{\alpha}$ corresponds to one input $x_{\alpha}$.
Calculating the network prior jointly for all inputs $x_{\alpha}$
is analogous to the replica calculation in physics \citep{ZinnJustin96,Hertz91}:
for each input $x_{\alpha}$ one considers a copy of the network with
the same network parameters $\theta$ shared across all of these replica.
Here, we showcase the derivation of $p(x_{\alpha}\vert y_{\beta})$
for a single input $x_{\alpha}$, dropping the sample index $\alpha$
for simplicity. The general case of $P$ inputs follows the same arguments
and is given in \prettyref{app:prior_multiple_inputs}.

The network prior is given by the probability of an output $y$ given
an input $x$ marginalized over the distribution of network parameters
\begin{equation}
p(y\vert x)=\int\mathrm{d}\theta\,p(y\vert x,\theta)\,p(\theta).\label{eq:network_prior_resnets}
\end{equation}
Given fixed network parameters $\theta$, the probability $p(y\vert x,\theta)$
is given by enforcing the network model with Dirac $\delta$-distributions
as{\small{}
\begin{align}
p(y\vert x,\theta) & =\int\mathrm{d}h^{(0)}\dots\int\mathrm{d}h^{(L)}\,\delta(h^{(0)}-W^{\text{in}}x-b^{\text{in}})\nonumber \\
 & \qquad\times\prod_{l=1}^{L}\delta(h^{(l)}-h^{(l-1)}-\rho W^{(l)}\phi(h^{(l-1)})-\rho b^{(l)})\nonumber \\
 & \qquad\times\delta(y-W^{\text{out}}\phi(h^{(L)})-b^{\text{out}}).\label{eq:prior_delta_enforced_resnets}
\end{align}
}{\small\par}

\subsubsection*{Marginalization over network parameters}

We marginalize over the network parameters as{\small{}
\begin{align}
p(y\vert x) & =\int\mathrm{d}h^{(0)}\dots\int\mathrm{d}h^{(L)}\,\langle\delta(h^{(0)}-W^{\text{in}}x-b^{\text{in}})\rangle_{\left\{ W^{\text{in}},\,b^{\text{in}}\right\} }\nonumber \\
 & \quad\times\prod_{l=1}^{L}\langle\delta(h^{(l)}-h^{(l-1)}-\rho W^{(l)}\phi^{(l-1)}-\rho b^{(l)})\rangle_{\left\{ W^{(l)},\,b^{(l)}\right\} }\nonumber \\
 & \quad\times\langle\delta(y-W^{\text{out}}\phi^{(L)}-b^{\text{out}})\rangle_{\left\{ W^{\text{out}},\,b^{\text{out}}\right\} },
\end{align}
}where $\left\langle \dots\right\rangle _{\{W,b\}}$ refers to the
expectation value over the statistics of weights $W$ and biases $b$
and we use the shorthand $\phi^{(l)}=\phi(h^{(l)})$. We rewrite the
Dirac $\delta$-distributions using its Fourier representation
\begin{align}
\delta(z) & =\prod_{k=1}^{N}\big\{\int_{-\infty}^{\infty}\frac{d\omega_{k}}{2\pi}\big\}\,e^{i\,\sum_{k=1}^{N}\omega_{k}\,z_{k}}\\
 & =\int\mathrm{d}\tilde{z}\,\exp\big(\tilde{z}^{\T}z\big)\label{eq:delta_fourier}
\end{align}
with scalar product $\tilde{z}^{\T}z=\sum_{i=1}^{N}\tilde{z}_{i}z_{i}$,
where we integrate along the imaginary axis $\int\mathrm{d}\tilde{z}=\prod_{k}\int_{i\mathbb{R}}\frac{\mathrm{d}\tilde{z}_{k}}{2\pi i}$.
Here, $\tilde{z}$ is referred to as the conjugate variable to $z$.
Writing the Fourier transform in complex variables is equivalent and
commonly used in related works \citep{Segadlo22_103401,seroussi23_908,Bordelon23_114009}.
This yields{\footnotesize{}
\begin{align}
 & p(y\vert x)\nonumber \\
 & =\int\mathrm{d}\tilde{y}\int\mathcal{D}\tilde{h}\int\mathcal{D}h\,\left\langle \exp\Big(\big(\tilde{h}^{(0)}\big)^{\T}(h^{(0)}-W^{\text{in}}x-b^{\text{in}})\Big)\right\rangle _{\left\{ W^{\text{in}},\,b^{\text{in}}\right\} }\nonumber \\
 & \quad\times\prod_{l=1}^{L}\left\langle \exp\Big(\big(\tilde{h}^{(l)}\big)^{\T}(h^{(l)}-h^{(l-1)}-\rho W^{(l)}\phi^{(l-1)}-\rho b^{(l)})\Big)\right\rangle _{\left\{ W^{(l)},\,b^{(l)}\right\} }\nonumber \\
 & \quad\times\left\langle \exp\Big(\tilde{y}^{\T}(y-W^{\text{out}}\phi^{(L)}-b^{\text{out}})\Big)\right\rangle _{\left\{ W^{\text{out}},\,b^{\text{out}}\right\} },
\end{align}
}where $\int\mathcal{D}h=\prod_{l=0}^{L}\int\mathrm{d}h^{(l)}$ and
${\int\mathcal{D}\tilde{h}=\prod_{l=0}^{L}\int\mathrm{d}\tilde{h}^{(l)}}$
for brevity.

Since the network parameters $\theta_{k}$ are independently distributed,
the integrals decouple and only integrals of the form $\int\mathrm{d}\theta_{k}\,p(\theta_{k})\,\exp\big(z\theta_{k}\big)$
appear, which can be solved exactly for $\theta_{k}\sim\mathcal{N}(0,\sigma^{2})$
yielding $\exp\big(\frac{1}{2}\sigma^{2}z^{2}\big)$. As an example,
we have{\footnotesize{}
\begin{align}
\bigg\langle\exp\Big(-\sum_{i,j}W_{ij}^{(l)}\,\rho\tilde{h}_{i}^{(l)}\phi_{j}^{(l-1)}\Big)\bigg\rangle_{W^{(l)}} & =\exp\Big(\frac{1}{2}\frac{\sigma_{w}^{2}}{N}\sum_{i,j}\Big(\rho\tilde{h}_{i}^{(l)}\phi_{j}^{(l-1)}\Big)^{2}\Big).
\end{align}
}Rewriting the resulting terms as {\small{}$\sum_{mn}\left[\tilde{h}_{m}\phi_{n}^{(l-1)}\right]^{2}=\tilde{h}^{\T}\tilde{h\,}\left[\phi^{(l-1)}\right]^{\T}\phi^{(l-1)}$},
we obtain the action $\mathcal{S}$ of the network prior
\begin{align}
p(y\vert x) & =\int\mathrm{d}\tilde{y}\int\mathcal{D}\tilde{h}\int\mathcal{D}h\,\exp\left(\mathcal{S}(y,\tilde{y},h,\tilde{h}\vert x)\right),
\end{align}
where we distinguish between the contributions of the readin layer,
the hidden layers of the network with residual connectivity, and the
readout layer
\begin{align}
\mathcal{S}(y,\tilde{y},h,\tilde{h}\vert x) & =\mathcal{S}_{\text{in}}(h^{(0)},\tilde{h}^{(0)}\vert x)+\mathcal{S}_{\text{net}}(h,\tilde{h})\nonumber \\
 & \qquad+\mathcal{S}_{\text{out}}(y,\tilde{y}\vert h^{(L)}).\label{eq:action_total}
\end{align}
These contributions are given by\begin{widetext}
\begin{align}
\mathcal{S}_{\text{in}}(h^{(0)},\tilde{h}^{(0)}\vert x) & \coloneqq\left[\tilde{h}^{(0)}\right]^{\T}h^{(0)}+\frac{1}{2}\frac{\sigma_{w,\,\text{in}}^{2}}{d_{\text{in}}}\,\left[\tilde{h}^{(0)}\right]^{\T}\tilde{h}^{(0)}\,x{}^{\T}x+\frac{1}{2}\sigma_{b,\,\text{in}}^{2}\left[\tilde{h}^{(0)}\right]^{\T}\tilde{h}^{(0)},\label{eq:action_in}\\
\mathcal{S}_{\text{net}}(h,\tilde{h}) & \coloneqq\sum_{l=1}^{L}\left[\tilde{h}^{(l)}\right]^{\T}\left[h^{(l)}-h^{(l-1)}\right]+\frac{1}{2}\rho^{2}\frac{\sigma_{w}^{2}}{N}\,\left[\tilde{h}^{(l)}\right]^{\T}\tilde{h}^{(l)}\,\left[\phi^{(l-1)}\right]^{\T}\phi^{(l-1)}+\frac{1}{2}\rho^{2}\sigma_{b}^{2}\left[\tilde{h}^{(l)}\right]^{\T}\tilde{h}^{(l)},\label{eq:action_net}\\
\mathcal{S}_{\text{out}}(y,\tilde{y}\vert h^{(L)}) & \coloneqq\tilde{y}^{\T}y+\frac{1}{2}\frac{\sigma_{w,\,\text{out}}^{2}}{N}\,\tilde{y}^{\T}\tilde{y}\,\left[\phi^{(L)}\right]^{\T}\phi^{(L)}+\frac{1}{2}\sigma_{b,\,\text{out}}^{2}\,\tilde{y}^{\T}\tilde{y}.\label{eq:action_out}
\end{align}
\end{widetext}In contrast to feed-forward networks \citep{Segadlo22_103401},
the conjugate variable $\tilde{h}^{(l)}$ of layer $l$ does not only
couple to the signal $h^{(l)}$ of layer $l$, but also to the signal
$h^{(l-1)}$ of the previous layer $l-1$. This coupling across layers
results from the skip connections in residual networks. The interdependence
between layers induced by the coupling prohibits the marginalization
over the intermediate signals $h^{(l)}$ in a direct manner as in
feed-forward networks.

\subsubsection*{Auxiliary variables}

Quadratic terms in $h$ and $\tilde{h}$ can be solved as Gaussian
integrals. However, in \eqref{eq:action_in}-\eqref{eq:action_out}
terms proportional to $\propto\left[\tilde{h}^{(l)}\right]^{\T}\tilde{h}^{(l)}\,\left[\phi^{(l-1)}\right]^{\T}\phi^{(l-1)}$
appear, which are at least quartic in $h$ and $\tilde{h}$ for linear
$\phi(h)=h$ and of even higher order for non-linear $\phi$. To treat
these terms, we introduce auxiliary variables\allowdisplaybreaks
\begin{align}
C^{(0)} & \coloneqq\frac{\sigma_{w,\,\text{in}}^{2}}{d_{\text{in}}}x{}^{\T}x+\sigma_{b,\,\text{in}}^{2},\label{eq:def_aux}\\
C^{(l)} & \coloneqq\rho^{2}\frac{\sigma_{w}^{2}}{N}\left[\phi^{(l-1)}\right]^{\T}\phi^{(l-1)}+\rho^{2}\sigma_{b}^{2}\quad l=1,\dots,L,\\
C^{(L+1)} & \coloneqq\frac{\sigma_{w,\,\text{out}}^{2}}{N}\left[\phi^{(L)}\right]^{\T}\phi^{(L)}+\sigma_{b,\,\text{out}}^{2}.
\end{align}
\allowdisplaybreaks[0]For wide networks $N\gg1$, we expect the
average $\frac{1}{N}\sum_{i=1}^{N}\left[\phi_{i}^{(l-1)}\right]^{2}$
to concentrate around its mean value. Based on this intuition, we
aim to rewrite the network prior $p(y\vert x)$ in terms of these
scalar variables. Note that, while the auxiliary variables are scalar
for a single input $x$, they become kernel matrices $C_{\alpha\beta}^{(l)}$
for multiple inputs with sample indices $\alpha,\,\beta$ (see \prettyref{app:prior_multiple_inputs}
for details).

We enforce the definitions of the auxiliary variables with Dirac $\delta$-distributions
as in \prettyref{eq:prior_delta_enforced_resnets}, e.g.{\footnotesize{}
\begin{align}
 & \delta\bigg(-NC^{(l)}+\rho^{2}\sigma_{w}^{2}\,\left[\phi^{(l-1)}\right]^{\T}\phi^{(l-1)}+N\rho^{2}\sigma_{b}^{2}\bigg)\label{eq:constraint_aux}\\
 & =\int_{i\R}\frac{\mathrm{d}\tilde{C}^{(l)}}{2\pi i}\,\exp\bigg(-N\tilde{C}^{(l)}C^{(l)}+\rho^{2}\sigma_{w}^{2}\,\tilde{C}^{(l)}\,\left[\phi^{(l-1)}\right]^{\T}\phi^{(l-1)}+N\rho^{2}\sigma_{b}^{2}\,\tilde{C}^{(l)}\bigg)\nonumber \\
 & =\int_{i\R}\frac{\mathrm{d}\tilde{C}^{(l)}}{2\pi i}\,\exp\bigg(-N\tilde{C}^{(l)}C^{(l)}+N\rho^{2}\sigma_{w}^{2}\,\tilde{C}^{(l)}\,\left[\phi_{i}^{(l-1)}\right]^{2}+N\rho^{2}\sigma_{b}^{2}\,\tilde{C}^{(l)}\bigg).\nonumber 
\end{align}
}In the last line, we used that the scalar variables $C$ and and
its conjugate variables $\tilde{C}$ only couple to sums of $\tilde{h}$
and $\phi(h)$ over all neuron indices, so that all components of
$h$, $\tilde{h}$, and thus also $\phi(h)$ are identically distributed.
Thus, we can rewrite the expression in scalar variables $h$ and $\tilde{h}$,
pulling out a factor $N$ in all terms. For the input layer, we denote
the ratio $d_{\text{in}}/N=\nu_{0}$ and set $\nu_{l}=1$ for $l>0$;
different network widths $N_{l}$ across layers $l$ can be considered
by setting $\nu_{l}=N_{l}/N$. Moving all integrals over scalar variables
$h$ and $\tilde{h}$ to the exponent, we can write the network prior
in terms of the kernels as
\begin{align}
p(y\vert x) & =\int\mathrm{d}\tilde{y}\left\langle \exp\left(\tilde{y}^{\T}y+\frac{1}{2}C^{(L+1)}\tilde{y}^{\T}\tilde{y}\right)\right\rangle _{C,\tilde{C}},
\end{align}
where the statistics of $C,\tilde{C}$ are given by $p(C,\tilde{C})\propto\exp\big(\mathcal{S}(C,\tilde{C})\big)$
with\begin{widetext}
\begin{align}
\mathcal{S}(C,\tilde{C}) & \coloneqq-N\sum_{l=1}^{L+1}\nu_{l}\,C^{(l)}\,\tilde{C}^{(l)}+N\mathcal{W}(\tilde{C}\vert C),\\
\mathcal{W}(\tilde{C}\vert C) & \coloneqq\ln\prod_{l=1}^{L}\int\mathrm{d}h^{(l)}\int\mathrm{d}\tilde{h}^{(l)}\,\exp\left(\tilde{h}^{(l)}\left[h^{(l)}-h^{(l-1)}\right]+\frac{1}{2}C^{(l)}\left[\tilde{h}^{(l)}\right]^{2}\right)\\
 & \qquad\qquad\times\exp\left(\tilde{C}^{(l)}\,\Big[\rho^{2}\sigma_{w}^{2}\,\phi^{(l-1)}\phi^{(l-1)}+\rho^{2}\sigma_{b}^{2}\Big]\right)\nonumber \\
 & \qquad\qquad\times\exp\left(\tilde{C}^{(L+1)}\,\bigg[\sigma_{w,\,\text{out}}^{2}\,\phi^{(L)}\phi^{(L)}+\sigma_{b,\,\text{out}}^{2}\bigg]\right)\nonumber \\
 & \qquad\qquad\times\int\mathrm{d}h^{(0)}\int\mathrm{d}\tilde{h}^{(0)}\exp\left(\tilde{h}^{(0)}h^{(0)}+\frac{1}{2}C^{(0)}\left[\tilde{h}^{(0)}\right]^{2}+\tilde{C}^{(0)}\,\Bigg[\frac{\sigma_{w,\,\text{in}}^{2}}{N}x{}^{\T}x+\nu_{0}\,\sigma_{b,\,\text{in}}^{2}\Bigg]\right).\nonumber 
\end{align}
\end{widetext}Note that the conjugate variables $\tilde{C}^{(l)}$
are not proper random variables, but will be integrated out later
to obtain the statistics of the auxiliary variables $C^{(l)}$.

\subsection{Saddle point approximation yields NNGP}

We obtain the NNGP kernel for residual networks in the limit of infinite
network width $N\rightarrow\infty$ as the saddle point of the action
$\mathcal{S}$ describing the network prior. Since the action $\mathcal{S}$
scales with the network width $N$, we can perform such a saddle point
approximation in the limit of infinite width $N\rightarrow\infty$
to evaluate integrals of the form
\begin{equation}
\int\mathcal{D}C\int\mathcal{D}\tilde{C}\,f(C,\tilde{C})\,\exp\big(\mathcal{S}(C,\tilde{C})\big)\overset{N\rightarrow\infty}{=}f(C_{*},\tilde{C}_{*}),
\end{equation}
where $C_{*}$ and $\tilde{C}_{*}$ are the saddle points of the action
$\mathcal{S}$.

We compute the saddle points using the conditions
\begin{equation}
\frac{\partial\mathcal{S}}{\partial C}=0,\;\frac{\partial\mathcal{S}}{\partial\tilde{C}}=0,
\end{equation}
and get
\begin{align}
C_{*}^{(l)} & =\begin{cases}
\frac{\sigma_{w,\,\text{in}}^{2}}{d_{\text{in}}}x^{\T}x+\sigma_{b,\,\text{in}}^{2} & l=0,\\
\rho^{2}\sigma_{w}^{2}\,\langle\phi^{(l-1)}\phi^{(l-1)}\rangle_{p}+\rho^{2}\sigma_{b}^{2} & 1\leq l\leq L,\\
\sigma_{w,\,\text{out}}^{2}\,\langle\phi^{(L)}\phi^{(L)}\rangle_{p}+\sigma_{b,\,\text{out}}^{2} & l=L+1,
\end{cases}\label{eq:res_kernel_self_cons}\\
\tilde{C}_{*}^{(l)} & =\begin{cases}
\frac{1}{2}\bigg\langle\bigg[\tilde{h}^{(l)}\bigg]^{2}\bigg\rangle_{p}=0 & l=0,\dots,L,\\
\frac{1}{2}\langle\tilde{y}^{2}\rangle_{p}=0 & l=L+1,
\end{cases}\label{eq:self_cons_C_tilde}
\end{align}
where{\footnotesize{}
\begin{align}
 & \langle\dots\rangle_{p}\nonumber \\
 & =\int\mathrm{d}h^{(0)}\int\mathrm{d}\tilde{h}^{(0)}\exp\left(\tilde{h}^{(0)}h^{(0)}+\frac{1}{2}C_{*}^{(0)}\left[\tilde{h}^{(0)}\right]^{2}\right)\\
 & \times\prod_{l=1}^{L}\int\mathrm{d}h^{(l)}\int\mathrm{d}\tilde{h}^{(l)}\,\dots\,\exp\left(\tilde{h}^{(l)}\left[h^{(l)}-h^{(l-1)}\right]+\frac{1}{2}C_{*}^{(l)}\left[\tilde{h}^{(l)}\right]^{2}\right).\nonumber 
\end{align}
}For brevity, we also include the input kernel $C_{*}^{(0)}=C^{(0)}$
here, which is fixed by the data as ${C^{(0)}=\sigma_{w,\,\text{in}}^{2}/d_{\text{in}}\,x{}^{\T}x+\sigma_{b,\,\text{in}}^{2}}$.
The expectation value over the auxiliary variables $\tilde{h}$ and
$\tilde{y}$ in \eqref{eq:self_cons_C_tilde} vanishes due to the
normalization the probability distribution, which we explain in \prettyref{app:saddle_point_approx}
in detail. In the following, we use the notation $\tilde{y}=\tilde{h}^{(L+1)}$
for brevity. The appearing expectation values are computed self-consistently
with respect to $C_{*}^{(l)}$.

By defining the residual $f^{(l)}\coloneqq h^{(l)}-h^{(l-1)}$ for
$1\leq l\leq L$ and $f^{(0)}\coloneqq h^{(0)}$, we rewrite the appearing
average as{\footnotesize{}
\begin{align}
 & \langle\dots\rangle_{p}\nonumber \\
 & =\int\mathrm{d}f^{(0)}\int\mathrm{d}\tilde{f}^{(0)}\exp\left(\tilde{f}^{(0)}f^{(0)}+\frac{1}{2}C_{*}^{(0)}\left[\tilde{f}^{(0)}\right]^{2}\right)\,\nonumber \\
 & \quad\times\prod_{l=1}^{L}\int\mathrm{d}f^{(l)}\int\mathrm{d}\tilde{h}^{(l)}\,\dots\,\exp\left(\tilde{h}^{(l)}f^{(l)}+\frac{1}{2}C_{*}^{(l)}\left[\tilde{h}^{(l)}\right]^{2}\right).\label{eq:def_mf_measure}
\end{align}
}We observe that the residuals $f^{(l)}$ for $0\leq l\leq L$ are
Gaussian distributed $f^{(l)}\sim\mathcal{N}(0,C_{*}^{(l)})$ with
zero mean and covariance $C_{*}^{(l)}$. Since the expectation values
in \eqref{eq:res_kernel_self_cons} have $h^{(l)}$ as an argument,
we would like to rewrite $\langle\dots\rangle_{p}$ with respect to
$h^{(l)}$. Since $h^{(l)}=\sum_{k=0}^{l}f^{(k)}$ and the residuals
$f^{(l)}$ are independent Gaussians, the signal $h^{(l)}$ is also
Gaussian distributed with covariance $K^{(l)}=\sum_{k=0}^{l}C_{*}^{(k)}$.

Thus, we can rewrite \eqref{eq:res_kernel_self_cons} as
\begin{align}
C_{*}^{(l)} & =\rho^{2}\sigma_{w}^{2}\,\langle\phi^{(l-1)}\phi^{(l-1)}\rangle_{\mathcal{N}(0,K^{(l-1)})}+\rho^{2}\sigma_{b}^{2},\label{eq:C_l}\\
K^{(l)} & =\begin{cases}
\frac{\sigma_{w,\,\text{in}}^{2}}{d_{\text{in}}}x^{\T}x+\sigma_{b,\,\text{in}}^{2} & l=0,\\
\sum_{k=0}^{l}C_{*}^{(k)} & 1\leq l\leq L,\\
\sigma_{w,\,\text{out}}^{2}\,\langle\phi^{(L)}\phi^{(L)}\rangle_{\mathcal{N}(0,K^{(L)})}+\sigma_{b,\,\text{out}}^{2} & l=L+1.
\end{cases}\label{eq:nngp_resnets_derived}
\end{align}
We recover the known NNGP result for the kernels as $K^{(l)}=K^{(l-1)}+C_{*}^{(l)}$
\citep{Huang20_33,Tirer22_921,Barzilai23}.

\subsection{Next-to-leading-order correction yields response function\label{subsec:leading_order_corr_resnets}}

The strength of the here employed field-theoretic formalism is that
finite-size corrections to the NNGP result in \eqref{eq:nngp_resnets_derived}
can be systematically calculated by expanding the action $\mathcal{S}$
in \eqref{eq:action_total} around its saddle point. In particular,
for finite-size networks, the residual kernels $C^{(l)}$ do not exactly
concentrate to the saddle point value but fluctuate around it. To
lowest-order, we describe these fluctuations as Gaussian. The Gaussian
fluctuations correspond to physical quantities such as the response
function: The response function of the network is defined as the linear
response of the covariance measured in layer $l$ to a perturbation
of the covariance in layer $m$. To define the perturbation, we introduce
the perturbed auxiliary variable \eqref{eq:def_aux} in the $m$-th
layer as $\bar{C}^{(m)}(\epsilon)\coloneqq C^{(m)}+\epsilon$ for
small $\epsilon$, which obey a perturbed probability measure $\text{(\ensuremath{\bar{C}},\ensuremath{\bar{\tilde{C}}})\ensuremath{\sim p_{m}(\epsilon)}}$.
Then, the response function is given by
\begin{align}
\Delta_{12}^{lm} & \coloneqq\lim_{\epsilon\to0}\,\frac{1}{\epsilon}\,\big(\langle\bar{C}^{(l)}\rangle\big|_{p_{m}(\epsilon)}-\langle\bar{C}^{(l)}\rangle\big|_{p_{m}(0)}\big).\label{eq:response_finite_diff}
\end{align}
Since the perturbation term $\epsilon$ appears in the constraint
\eqref{eq:constraint_aux} in the form 
\[
\ldots\exp\bigg(N\tilde{C}^{(m)}\big[-\bar{C}^{(m)}+\epsilon+N\rho^{2}\sigma_{w}^{2}\,\left[\phi_{i}^{(l-1)}\right]^{2}+N\rho^{2}\sigma_{b}^{2}\,\big]\bigg),
\]
 this definition is equivalent to the correlator
\begin{align}
\Delta_{12}^{lm} & =N\,\big\langle C^{(l)}\tilde{C}^{(m)}\big\rangle,\label{eq:Delta_as_correlator}
\end{align}
as one may regard $\epsilon$ as a source term for $N\,\tilde{C}^{(m)}$.
The response function measures the network's sensitivity to different
signals and thus is a measure for signal propagation. In turn, signal
propagation in networks is a key indicator of network trainability
and performance \citep{Schoenholz17_01232}.

We obtain the Gaussian fluctuations of the kernels by computing the
Hessian $S^{(2)}\vert_{(C_{\ast},\tilde{C}_{\ast})}$ of the action
$\mathcal{S}$ at the saddle point
\begin{align}
\mathcal{S}^{(2)}\vert_{(C_{\ast},\tilde{C}_{\ast})} & =\left(\begin{array}{cc}
\frac{\partial^{2}}{\partial C^{2}}\mathcal{S} & \frac{\partial^{2}}{\partial C\,\partial\tilde{C}}\mathcal{S}\\
\frac{\partial^{2}}{\partial\tilde{C}\,\partial C}\mathcal{S} & \frac{\partial^{2}}{\partial\tilde{C}^{2}}\mathcal{S}
\end{array}\right)\rvert_{(C_{\ast},\tilde{C}_{\ast})}\eqqcolon\left(\begin{array}{cc}
\mathcal{S}_{11} & \mathcal{S}_{12}\\
\mathcal{S}_{21} & \mathcal{S}_{22}
\end{array}\right).
\end{align}
Due to the evaluation at the saddle point, all following expectation
values are with respect to the measure $\langle\dots\rangle_{p}$
defined in \eqref{eq:def_mf_measure}. In the following, we drop $\vert_{(C_{\ast},\tilde{C}_{\ast})}$
for notational brevity. The diagonal blocks of the Hessian are given
by
\begin{align}
\frac{\partial^{2}}{\partial C^{(l)}\partial C^{(k)}}\mathcal{S} & =\frac{1}{4}\big\langle\tilde{h}^{(l)}\tilde{h}^{(l)},\tilde{h}^{(k)}\tilde{h}^{(k)}\big\rangle_{p}^{c}=0,\label{eq:hessian_virtual}\\
\frac{\partial^{2}}{\partial\tilde{C}^{(l)}\partial\tilde{C}^{(k)}}\mathcal{S} & =N\sigma_{w}^{4}\,1_{l>0}1_{k>0}\,\langle\phi^{(l-1)}\phi^{(l-1)},\phi^{(k-1)}\phi^{(k-1)}\rangle_{p}^{c}\nonumber \\
 & \qquad\times\begin{cases}
\rho^{4} & k,l\neq L+1,\\
\rho^{2} & k\neq l=L+1\vee l\neq k=L+1,\\
1 & \text{else},
\end{cases}
\end{align}
where $1_{l>0}$ denotes the indicator function. We write $\langle\dots\rangle^{c}$
for connected correlations defined as
\begin{align}
 & \langle z^{(l)}z^{(l)},z^{(k)}z^{(k)}\rangle_{p}^{c}\\
 & =\langle z^{(l)}z^{(l)}z^{(k)}z^{(k)}\rangle_{p}-\langle z^{(l)}z^{(l)}\rangle_{p}\,\langle z^{(k)}z^{(k)}\rangle_{p}.\nonumber 
\end{align}
The average over the auxiliary variables $\tilde{h}$ in \eqref{eq:hessian_virtual}
vanishes due to the normalization of the probability distribution,
which we explain in \prettyref{app:saddle_point_approx} in detail.
For the off-diagonal terms, we have
\begin{align}
 & \frac{\partial^{2}}{\partial C^{(l)}\partial\tilde{C}^{(k)}}\mathcal{S}\label{eq:offdiag_hess}\\
 & =-N\nu_{l}\delta_{kl}\nonumber \\
 & \quad+N\,1_{k>0}\,\sigma_{w}^{2}\langle\left[\phi^{\prime,(k-1)}\right]^{2}+\phi^{\prime\prime,(k-1)}\phi^{(k-1)}\rangle_{\mathcal{N}(0,K^{(k-1)})}\,1_{k>l}\nonumber \\
 & \qquad\times\begin{cases}
\rho^{2} & k\leq L\\
1 & k=L+1
\end{cases}\,,\nonumber 
\end{align}
where we used Price's theorem for a one-dimensional variable \citep{PapoulisProb4th}
\begin{equation}
\partial_{A}\langle\phi^{2}(h)\rangle_{h\sim\mathcal{N}(0,A)}=\langle\partial_{h}^{2}\phi^{2}(h)\rangle_{h\sim\mathcal{N}(0,A)}\label{eq:price_onedim}
\end{equation}
to compute the derivative of the expectation value by the covariance
(see \prettyref{app:prices_theorem} for details). The condition
$k>l$ enforced by the indicator function $1_{k>l}$ results from
a term $\partial_{C^{(l)}}K^{(k-1)}$ appearing in the derivative,
because the network kernel $K^{(k-1)}$ only depends on the residual
kernels $C^{(l)}$ with $l<k$.

We obtain the Gaussian fluctuations of the variables $C^{(l)}$ and
$\tilde{C}^{(l)}$ by taking the negative inverse of the Hessian,
also called the propagator $\mathcal{G}$ in field theory
\begin{align}
\mathcal{G} & \coloneqq-(\mathcal{S}^{(2)})^{-1}\eqqcolon\left(\begin{array}{cc}
\mathcal{G}_{11} & \mathcal{G}_{12}\\
\mathcal{G}_{21} & \mathcal{G}_{22}
\end{array}\right).
\end{align}
By using the block structure and the fact that $\mathcal{S}_{11}=0$,
we have
\begin{align}
\mathcal{G}_{11} & =\mathcal{G}_{12}\,\mathcal{S}_{22}\,\mathcal{G}_{21},\label{eq:delta11}\\
\mathcal{G}_{12} & =-\mathcal{S}_{21}^{-1},\\
\mathcal{G}_{22} & =0.
\end{align}
Since the off-diagonal block matrix $\mathcal{S}_{21}$ is a lower
triangular matrix, its inverse can be computed using forward propagation
\begin{align}
 & \mathcal{G}_{12}^{lm}\\
 & =N^{-1}\nu_{l}^{-1}\delta_{lm}\nonumber \\
 & \quad+1_{L\geq l>0}\,\rho^{2}\sigma_{w}^{2}\langle\left[\phi^{\prime,(l-1)}\right]^{2}+\phi^{\prime\prime,(l-1)}\phi^{(l-1)}\rangle_{\mathcal{N}(0,K^{(l-1)})}\,\sum_{k=0}^{l-1}\mathcal{G}_{12}^{km}\nonumber \\
 & \quad+\delta_{l,L+1}\,\sigma_{w}^{2}\langle\left[\phi^{\prime,(l-1)}\right]^{2}+\phi^{\prime\prime,(l-1)}\phi^{(l-1)}\rangle_{\mathcal{N}(0,K^{(l-1)})}\,\sum_{k=0}^{l-1}\mathcal{G}_{12}^{km},\nonumber 
\end{align}
yielding the response function ${\Delta_{12}^{lm}=N\langle C^{(l)}\tilde{C}^{(m)}\rangle}=N\mathcal{G}_{12}^{lm}$
as a $\mathcal{O}(N^{-1})$ correction to the NNGP result. Due to
the residual architecture, any response can only propagate forward
in the network, which is reflected in the term $1_{k>l}$ in \eqref{eq:offdiag_hess}.
The kernel $K^{(l)}$ in layer $l$ depends explicitly not only on
the kernel of the previous layer as for FFNets but on all preceding
layers. This property directly results from the skip connections in
residual networks. Thereby, the response function in layer $l$ contains
information from all preceding layers, such that information can propagate
to deeper network layers.

We are ultimately interested in the response with respect to the network
input, hence we define 
\begin{equation}
\eta^{(l)}\coloneqq\Delta_{12}^{l\,0}.
\end{equation}
For the response of the residual kernels $C^{(l)}$ in all intermediate
layers $0<l<L+1$, we then have{\footnotesize{}
\begin{equation}
\eta^{(l)}=\rho^{2}\sigma_{w}^{2}\langle\left[\phi^{\prime,(k-1)}\right]^{2}+\phi^{\prime\prime,(k-1)}\phi^{(k-1)}\rangle_{\mathcal{N}(0,K^{(l-1)})}\,\sum_{k=0}^{l-1}\eta^{(k)}.\label{eq:eta}
\end{equation}
}In particular, it is $\eta^{(0)}=N/d_{\text{in}}$. Due to their
additive nature, the response function $\chi^{(l)}$ of the kernels
$K^{(l)}$ is given by 
\begin{align}
\chi^{(l)} & \coloneqq\langle K^{(l)}\,\tilde{C}^{(0)}\rangle\\
 & =\sum_{k=0}^{l}\eta^{(k)}.
\end{align}
Finally, the output response $\chi^{\text{out}}$ is given by{\footnotesize{}
\begin{align}
\chi^{\text{out}} & \coloneqq\langle K^{(L+1)}\,\tilde{C}^{(0)}\rangle\label{eq:output_response}\\
 & =\sigma_{w,\,\text{out}}^{2}\,\langle\left[\phi^{(L)}\right]^{\prime}\left[\phi^{(L)}\right]^{\prime}+\left[\phi^{(L)}\right]^{\prime\prime}\left[\phi^{(L)}\right]\rangle_{\mathcal{N}(0,K^{(L)})}\,\sum_{k=0}^{L}\,\eta^{(k)}.
\end{align}
}In addition, one can also compute the fluctuation corrections from
the Hessian via \eqref{eq:delta11}. For details, see \prettyref{app:leading_corr_multidim}.

\begin{figure}
\centering{}\includegraphics{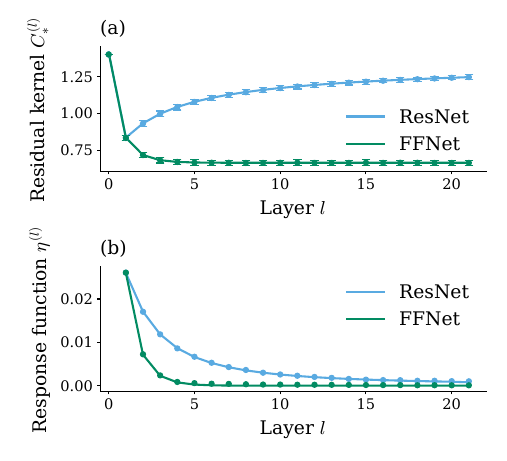}\caption{\foreignlanguage{american}{Residual kernels $C_{*}^{(l)}$ (a) and the respective response function
$\eta^{(l)}$ (b) in ResNets (blue) compared to FFNets (green). In
(a) error bars indicate standard error of the mean obtained from simulation
over $10^{3}$ network initializations, solid curves show theory values
from \eqref{eq:C_l}. In (b) dots represent simulations over $10^{2}$
input samples and $10^{3}$ network initializations, solid curves
show theory values from \eqref{eq:eta}. Errors are of order $10^{-5}$
and therefore not shown. ResNets exhibit a slower decay over layers
$l$ compared to FFNets. Other parameters: $\sigma_{w,\,\text{in}}^{2}=\sigma_{w}^{2}=\sigma_{w,\,\text{out}}^{2}=1.2,$
$\sigma_{w,\,\text{in}}^{2}=\sigma_{w}^{2}=\sigma_{w,\,\text{out}}^{2}=1.2,\,\sigma_{b,\,\text{in}}^{2}=\sigma_{b}^{2}=\sigma_{b,\,\text{out}}^{2}=0.2,$
$d_{\text{in}}=d_{\text{out}}=100,\,N=500,\,\rho=1$, $\phi=\text{erf}$.\label{fig:sim_vs_theory}}}
\end{figure}

In \prettyref{fig:sim_vs_theory} we compare the behavior of the residual
kernels $C_{*}^{(l)}$ and the response function $\eta^{(l)}$ between
FFNets and ResNets. While the residual kernels $C_{*}^{(l)}$ in FFNets
decay to zero as a function of depth, they approach a value larger
than zero in ResNets due to the accumulation of variance across layers.
Similarly, while the response function in FFNets decays exponentially
to zero, it decays much slower in ResNets and approaches zero only
asymptotically (see \prettyref{fig:decay_response} in \prettyref{app:decay_response}).
The latter observation matches previous results by \citep{Yang17_30}
based on the convergence rate of the kernels; however, we here derive
the response function that explicitly measures the dependence on the
input kernel.

\subsection{Relation to linear response theory}

In the previous section, we derived the response function of the network
as the first-order approximation in $\mathcal{O}\big(N^{-1}\big)$
from a systematic field-theoretic calculation. The field-theoretic
approach has the advantage of providing a framework in which we can
systematically compute such correction terms to different orders;
to provide more intuition on the response function we here discuss
its relation to linear response theory.

Consider a change of the input kernel given by ${K^{(0)}=\langle K^{(0)}\rangle+\delta K^{(0)}}$.
For small perturbations $\delta K^{(0)}$, we can ask how kernels
in subsequent layers are affected by this perturbation. To linear
order in $\delta K^{(0)}$, we expand the residual kernel as
\begin{equation}
C_{*}^{(l)}=C_{*}^{(l)}\rvert_{\langle K^{(0)}\rangle}+\frac{\partial C^{(l)}}{\partial K^{(0)}}\rvert_{\langle K^{(0)}\rangle}\,\delta K^{(0)}+\mathcal{O}\big(\delta^{2}\big).
\end{equation}
The first-order Taylor term accounts for the effect of the perturbation
and corresponds to the response function of the residual kernels $C_{\ast}^{(l)}$
as
\begin{align}
\eta^{(l)} & =\frac{\partial C^{(l)}}{\partial K^{(0)}}\rvert_{\langle K^{(0)}\rangle}\\
 & =\rho^{2}\sigma_{w}^{2}\frac{\partial}{\partial K^{(l-1)}}\langle\phi^{(l-1)}\phi^{(l-1)}\rangle_{\mathcal{N}(0,K^{(l-1)})}\,\frac{\partial K^{(l-1)}}{\partial K^{(0)}}\rvert_{\langle K^{(0)}\rangle}\\
 & =\rho^{2}\sigma_{w}^{2}\langle\left[\phi^{\prime,(l-1)}\right]^{2}+\phi^{\prime\prime,(l-1)}\phi^{(l-1)}\rangle_{\mathcal{N}(0,K^{(l-1)})}\sum_{k=0}^{l-1}\eta^{(k)},
\end{align}
where we used Price's theorem \citep{PapoulisProb4th} as in \eqref{eq:price_onedim}
from the second to the third line and in the last line that
\begin{equation}
\frac{\partial K^{(l-1)}}{\partial K^{(0)}}\rvert_{\langle K^{(0)}\rangle}=\sum_{k=0}^{l-1}\frac{\partial C^{(k)}}{\partial K^{(0)}}\rvert_{\langle K^{(0)}\rangle}=\sum_{k=0}^{l-1}\eta^{(k)}.
\end{equation}
The expectation value of the derivatives ${\langle\left[\phi^{\prime,(l-1)}\right]^{2}+\phi^{\prime\prime,(l-1)}\phi^{(l-1)}\rangle_{\mathcal{N}(0,K^{(l-1)})}}$
measures how the perturbation of the kernel $K^{(l-1)}$ affects the
residual kernel $C_{*}^{(l)}$ in the subsequent layer $l$. It gets
multiplied by the accumulated perturbations of all previous layers,
as one expects intuitively due to the skip connections in residual
networks. The expression for the linear response of the kernels $K^{(l)}$
follows directly from its definition
\begin{equation}
\chi^{(l)}=\frac{\partial K^{(l)}}{\partial K^{(0)}}\rvert_{\langle K^{(0)}\rangle}=\frac{\partial}{\partial K^{(0)}}\sum_{k=0}^{l}C_{*}^{(k)}\rvert_{\langle K^{(0)}\rangle}=\sum_{k=0}^{l}\eta^{(l)}.
\end{equation}
While expressions for the response function can be computed using
linear response theory, the field-theoretic formalism in \prettyref{subsec:leading_order_corr_resnets}
formally shows that the response function is the leading order $\mathcal{O}\big(N^{-1}\big)$
finite-size correction to the NNGP result.

\section{Signal propagation and optimal scaling in residual networks\label{sec:signal_prop_scaling}}

Next we apply the field-theoretic framework derived above to study
signal propagation in ResNets. The sensitivity of signal propagation
to different inputs can be measured by the response function \citep{Schoenholz17_01232}.
Good signal propagation is linked to improved trainability and thus
higher generalization performance of trained networks \citep{Schoenholz17_01232,Yang17_30}.
We first focus on the behavior of the diagonal elements $K_{\alpha\alpha}^{(l)}$
of the network kernels on the same sample $x_{\alpha}$ to obtain
an intuition for the signal propagation in residual networks and then
discuss how their behavior extends to off-diagonal elements $K_{\alpha\beta}^{(l)}$
of the network kernels for different samples $x_{\alpha}\neq x_{\beta}$.
We drop the sample index where only a single sample is considered
and write explicitly where multiple samples are considered.

\subsection{Optimal scaling of the residual branch}

\begin{figure}
\includegraphics{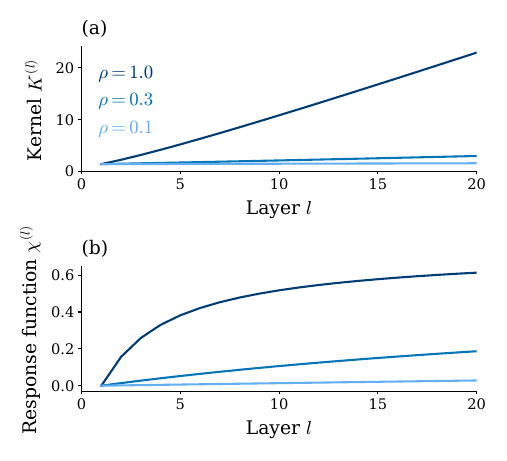}\caption{\foreignlanguage{american}{Dependence of (a) kernels $K^{(l)}$ and (b) the respective response
function $\chi^{(l)}$ on the residual scaling parameter $\rho$.
The residual scaling takes values $\rho\in[1.0,\,0.3,\,0.1]$ (from
dark to light). The residual scaling parameter $\rho$ governs the
rate of increase in both quantities. Other parameters: $\sigma_{w,\,\text{in}}^{2}=\sigma_{w}^{2}=\sigma_{w,\,\text{out}}^{2}=1.2$,
$\sigma_{b,\,\text{in}}^{2}=\sigma_{b}^{2}=\sigma_{b,\,\text{out}}^{2}=0.2$,
$d_{\text{in}}=d_{\text{out}}=100$, $N=500$, $\phi=\text{erf}$.\label{fig:summed_kernels}}}
\end{figure}

We start by studying the effect of the residual scaling parameter
$\rho$ on the kernels $K^{(l)}$ and response function $\chi^{(l)}$
of layer $l$ as these quantities describe the distribution of the
signal $h^{(l)}$. Since $\rho^{2}$ scales the residual kernels $C_{*}^{(l)}$
in \eqref{eq:C_l}, which are being summed to obtain $K^{(l)}$, the
residual scaling governs the rate of increase of $K^{(l)}$ across
layers (see \prettyref{fig:summed_kernels}(a)). The response function
$\chi^{(l)}$ exhibits the same scaling and thus behavior (see \prettyref{fig:summed_kernels}(b)).

\begin{figure*}
\includegraphics{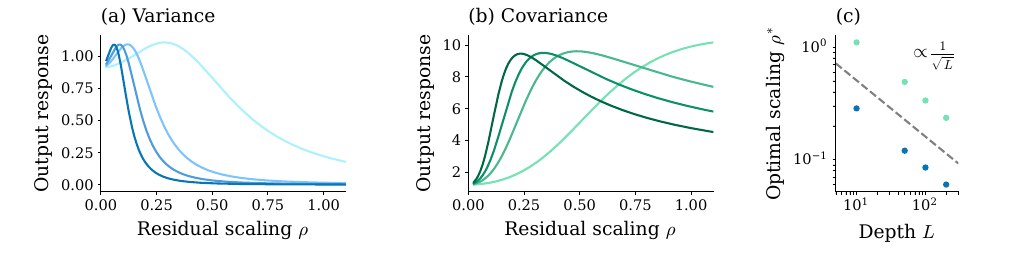}\caption{Optimal scaling of the residual branch. Output response $\chi^{\text{out}}$
for (a) diagonal and (b) off-diagonal elements of the network kernel
$K_{\alpha\beta}^{(l)}$. Different curves correspond to different
network depths $L\in[10,\,50,\,100,\,200]$ (light to dark). All curves
exhibit a unique maximum; the residual scaling values $\rho^{\ast}$
with largest response concentrate with increasing depth. (c) Optimal
residual scaling $\rho^{\ast}=\mathrm{argmax}(\chi^{\text{out}})$
for diagonal (blue) and off-diagonal (green) elements of the network
kernel $K_{\alpha\beta}^{(l)}$. In both cases, these scale with $1/\sqrt{L}$
(gray). Other parameters: input kernel $K^{(0)}=\left(\begin{smallmatrix}0.05 & 0.03\\0.03 & 0.05\end{smallmatrix}\right)$,
\foreignlanguage{american}{$\sigma_{w}^{2}=1.25$, $\sigma_{b}^{2}=0.05$,
$\ensuremath{d_{\text{in}}=d_{\text{out}}=100}$, $N=500$, $\phi=\text{erf}$.\label{fig:output_response_layers}}}
\end{figure*}

The dependence of the residual kernels $C_{*}^{(l)}$ on the residual
scaling $\rho$ transfers to the response function; due to the output
layer being a non-linear feed-forward layer the output response shown
in \prettyref{fig:output_response_layers}(a)-(b) exhibits a complex
non-linear behavior with a unique maximum $\rho^{*}$. The shape
of the response function and thus the optimal value $\rho^{\ast}$
depend on the network depth $L$, shifting to smaller values $\rho^{\ast}$
with larger depth. However, we observe an antagonistic effect: the
depth dependence becomes weaker for deeper networks. The optimal value
$\rho^{\ast}$ lies between $[0.1,\,0.3]$, as found empirically in
previous works \citep{Szegedy17_4278,Zhang19_4285}.

Due to the recursive nature of the non-linear Eqs. \eqref{eq:eta}-\eqref{eq:output_response}
for the response function, we cannot determine the optimal value $\rho^{\ast}$
analytically. However, for the variance $K^{(l)}$ we can make an
intuitive argument regarding the signal propagation in the network:
For deeper networks, the kernels $K^{(l)}$ in \prettyref{eq:nngp_resnets_derived}
grow continuously, so that the signal $h^{(l)}$ leaves the dynamic
range $\mathcal{V}$ of the activation function $\phi$ (see \prettyref{fig:graphical_abstract}(b)).
In consequence, part of the signal $h^{(L)}$ is lost in the readout
layer, reducing the output response $\chi^{\text{out}}$ to changing
inputs. The magnitude of the network kernels $K^{(l)}$ depends on
$\rho^{2}$, so that a smaller residual scaling leads to a less rapid
growth of the kernels $K^{(l)}$ and the signal $h^{(L)}$ stays in
the dynamic range $\mathcal{V}$. For very small scalings $\rho$,
the residual branch is suppressed and the network reduces to a single
hidden-layer perceptron (see \prettyref{fig:graphical_abstract}(c)).

Based on this intuition, we obtain an approximate expression for the
optimal scaling $\rho^{\ast}$: We assume that the signal $h^{(l)}$
stays in the dynamic range $\mathcal{V}$ of the activation function
so that $\phi(h^{(l)})\approx\phi^{\prime}(0)\,h^{(l)}$, where $\phi^{\prime}(0)$
accounts for the slope of the activation function at its origin. The
residual kernel then simplifies to 
\begin{equation}
C_{*}^{(l)}=\rho^{2}\sigma_{w}^{2}\,\phi^{\prime}(0)^{2}\,\sum_{k=0}^{l-1}C_{*}^{(k)}+\rho^{2}\sigma_{b}^{2},
\end{equation}
and hence we get
\begin{equation}
C_{*}^{(l)}=C_{*}^{(l-1)}+\rho^{2}\sigma_{w}^{2}\,\phi^{\prime}(0)^{2}\,C_{*}^{(l-1)}.
\end{equation}
Solving this recursion, we obtain
\begin{equation}
C_{*}^{(l)}=(1+\rho^{2}\sigma_{w}^{2}\,\phi^{\prime}(0)^{2})^{l-1}\,(\rho^{2}\sigma_{w}^{2}\,\phi^{\prime}(0)^{2}\,K^{(0)}+\rho^{2}\sigma_{b}^{2}).
\end{equation}
Using the sum of the first $L+1$ terms of the geometric series and
$C^{(0)}=K^{(0)}$ per definition yields
\begin{align}
K^{(L)} & =\sum_{k=0}^{L}C^{(k)}\\
 & =(1+\rho^{2}\sigma_{w}^{2}\,\phi^{\prime}(0)^{2})^{L}K^{(0)}\nonumber \\
 & \quad+\frac{\sigma_{b}^{2}}{\phi^{\prime}(0)^{2}\sigma_{w}^{2}}\left((1+\rho^{2}\sigma_{w}^{2}\,\phi^{\prime}(0)^{2})^{L}-1\right).\label{eq:output_kernel}
\end{align}
Assuming the $1\sigma$-range of the distribution to stay within the
dynamic range $\mathcal{V}$ for a point-symmetric activation function
$\phi$, we set $\mathcal{V}/2=\sqrt{K^{(L)}}$ and obtain for the
optimal scaling parameter
\begin{equation}
\rho^{\ast}\approx\frac{1}{\sigma_{w}\,\phi^{\prime}(0)}\sqrt{\left(\frac{\sigma_{w}^{2}\,\phi^{\prime}(0)^{2}\,\left(\nicefrac{\mathcal{V}}{2}\right)^{2}+\sigma_{b}^{2}}{\sigma_{w}^{2}\,\phi^{\prime}(0)^{2}\,K^{(0)}+\sigma_{b}^{2}}\right)^{\frac{1}{L}}-1}.\label{eq:opt_scaling}
\end{equation}
The assumption $\mathcal{V}/2=\sqrt{K^{(L)}}$ is only an estimate;
multiple $\sigma$-ranges could be required for optimal signal propagation.
Alternatively, we derive this condition from a maximum-entropy argument
for the signal distribution (see \prettyref{app:bukva_max_ent}).
A similar-maximum entropy argument has been used by \citet{Bukva23_arxiv}
to study trainability of feed-forward networks. Note that due to the
assumptions made in deriving this expression, it cannot fully capture
the behavior of the signal when it reaches the non-linear part of
the activation function. However, this is solely a limitation of this
particular ansatz to obtain an analytically tractable expression for
$\rho^{\ast}$; the response function itself depends in a non-linear
manner on the hyperparameters of the network and on the non-linear
activation function of the network.

\subsection{Depth scaling dominates optimal scaling}

\begin{figure}
\includegraphics{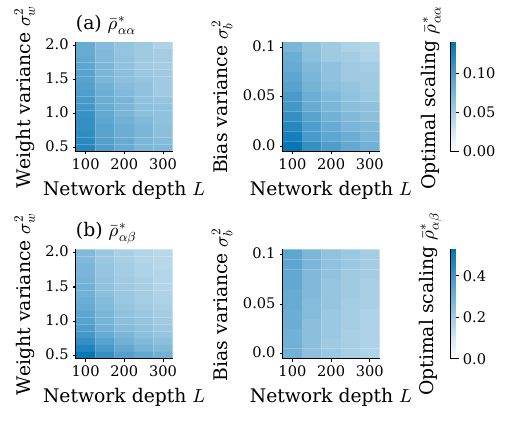}\caption{\foreignlanguage{american}{Optimal scalings depend strongly on network depth but weakly on other
hyperparameters. We illustrate the weak dependence on the weight variance
$\sigma_{w}^{2}$ and bias variance $\sigma_{b}^{2}$ relative to
the network depth $L$ for CIFAR-10 for both (a) variances and (b)
covariances; samples are either dogs or airplanes. We measure the
scaling with maximal output response averaged over all diagonal or
all off-diagonal elements of the covariance, $\bar{\rho}_{\alpha\alpha}^{\ast}=\frac{1}{N}\sum_{\alpha}\mathrm{argmax}(\chi_{\alpha\alpha}^{\text{out}})$
or $\bar{\rho}_{\alpha\beta}^{\ast}=\frac{1}{N(N-1)}\sum_{\alpha\protect\neq\beta}\mathrm{argmax}(\chi_{\alpha\beta}^{\text{out}})$.
Other parameters: \foreignlanguage{english}{data set size $P=20$,
input scale $K^{(0)}=0.05$, }$\sigma_{w}^{2}=1.25\,,\ensuremath{\sigma_{b}^{2}=0.05,\,d_{\text{in}}=d_{\text{out}}=100,\,N=500}$,
$\phi=\text{erf}$. \label{fig:param_scan}}}
\end{figure}
At large depth $L$, the expression for optimal scaling in \eqref{eq:opt_scaling}
is dominated by the appearing $L$-th root and can be written as
\begin{equation}
\rho^{\ast}\approx\sqrt{\frac{1}{L}}\sqrt{\frac{1}{\sigma_{w}^{2}\,\phi^{\prime}(0)^{2}}\log\left(\frac{\sigma_{w}^{2}\,\phi^{\prime}(0)^{2}\,\left(\nicefrac{\mathcal{V}}{2}\right)^{2}+\sigma_{b}^{2}}{\sigma_{w}^{2}\,\phi^{\prime}(0)^{2}\,K^{(0)}+\sigma_{b}^{2}}\right)},\label{eq:alpha_approx}
\end{equation}
neglecting terms of order $\mathcal{O}\left(L^{-1}\right)$. Thus,
we obtain the proportionality of the optimal scaling value with $\propto1/\sqrt{L}$
in \prettyref{fig:output_response_layers}(c); this scaling has been
reported based on different approaches in earlier works \citep{Hayou21_1324,Hayou21_iclr,Zhang22_3359}.
In contrast to those works, we here also obtain the dependence on
other hyperparameters of the network. While the scaling with $1/\sqrt{L}$
dominates, there is a weak dependence on other hyperparameters due
to the appearing logarithm as shown in \prettyref{fig:param_scan}.
This weak dependence on network hyperparameters but strong dependence
on network depth offers an explanation for the widespread success
of the $1/\sqrt{L}$ scaling across different architectures \citep{Bordelon24_iclr}.

\subsection{Behavior across full data set}

While the above considerations apply to the diagonal elements $K_{\alpha\alpha}^{(l)}$
of the network kernels, i.e. for the statistics of a single sample
$x_{\alpha}$, we require efficient signal propagation for all samples
of a data set of size $P$. The joint statistics of multiple data
samples cannot be analyzed with the simplified intuition of saturation
arguments and variances. The non-approximated field-theoretic result,
however, still yields the behavior of the full covariances. The off-diagonal
elements do not get a contribution from the term involving the second
derivative in \eqref{eq:opt_scaling}, which is negative for $\phi=\text{erf}$
and other saturating activation functions. In consequence, the response
function for off-diagonal elements is larger and admits a larger residual
scaling $\rho^{\ast}$.
\begin{figure*}
\includegraphics{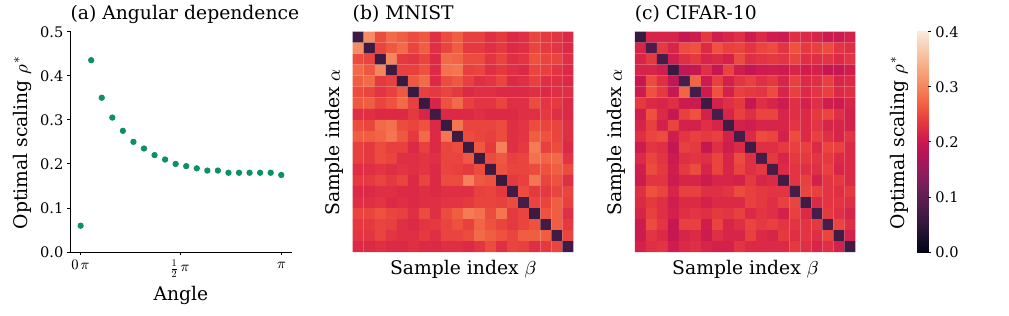}\caption{\foreignlanguage{american}{Dependence of optimal scaling $\rho^{\ast}$ on sample variability.
(a) \foreignlanguage{english}{Dependence of optimal scaling on the
angle between normalized synthetic data samples. On common, real-world
data sets such as (b) MNIST and (c) CIFAR-10, optimal scalings $\rho^{\ast}$
for off-diagonal elements of the kernels (samples sorted by classes
only) yield similar values across data samples. For MNIST we use samples
from digit $0$ and digit $3$, and for CIFAR-10 samples are either
dogs or airplanes.} \foreignlanguage{english}{Other parameters: data
set size $P=20$, input scale $K^{(0)}=0.05$, $\sigma_{w}^{2}=1.25$,
}$\sigma_{b}^{2}=0.05$, $\ensuremath{d_{\text{in}}=d_{\text{out}}=100}$,
$N=500$, $L=200$, $\phi=\text{erf}$.\label{fig:opt_response_full_kernels}}}
\end{figure*}

To investigate the dependence of the optimal scaling with maximal
output response $\rho^{\ast}(K_{\alpha\beta})=\mathrm{argmax}(\chi_{\alpha\beta}^{\text{out}})$
on differences in data samples, we generate samples of unit length
and encode sample differences by angles that have an equal spacing
in the range of $[0,2\pi]$ by steps of $2\pi/P$. In \prettyref{fig:opt_response_full_kernels}(a),
we observe a noticeable angular dependence of the optimal scaling.
For common data sets such as MNIST and CIFAR-10, we find that optimal
scalings $\rho^{\ast}(K_{\alpha\beta})$ for off-diagonal elements
of the kernels behave rather homogeneously, even though samples stem
from two different classes. Thus, the average scalings $\bar{\rho}^{\ast}=\frac{1}{N(N-1)}\sum_{\alpha\neq\beta}\mathrm{argmax}(\chi_{\alpha\beta}^{\text{out}})$
for off-diagonal elements of each data set provide a good indicator
for suitable values.

Further, we find the same $1/\sqrt{L}$ scaling of variances for the
covariances in \prettyref{fig:output_response_layers}(c). Apart from
this strong dependence on network depth, there is again only a weak
dependence of the optimal scaling on other hyperparameters (see \prettyref{fig:param_scan}(b)).

In summary, the response function, which naturally arises in a field-theoretic
formulation of residual networks, predicts the dependence of optimal
scaling $\rho^{\ast}$ on all network hyperparameters and data statistics
and thus provides a theoretical explanation for the empirically well-tested
$1/\sqrt{L}$ scaling.

\subsection{Behavior in trained networks}

We obtain the response function as a next-to-leading-order term to
the NNGP from the network prior. Thus, its form strictly only holds
for networks at initialization. \citet{Schoenholz17_01232} show for
feed-forward networks that signal propagation at initialization is
a useful indicator of network performance at training time. We therefore
study in this section whether the properties of the response function
in residual networks are robust at training time.

For the chosen scaling of weights $w\propto1/\sqrt{N}$, the prior
distribution of weights is known to change only very mildly due to
training, a result often termed ``lazy learning'' \citep{Chizat19_neurips}.
One therefore expects that predictions of the theory maintain their
validity, at least approximately. To check this, we train networks
on a binary classification between digits $0$ and $3$ of MNIST using
Langevin gradient descent \citep{Welling11_681,Stephan17_1,Naveh21_064301}
until convergence. The stationary weight distribution can alternatively
be regarded as drawn from the Bayesian posterior.

\begin{figure*}
\includegraphics{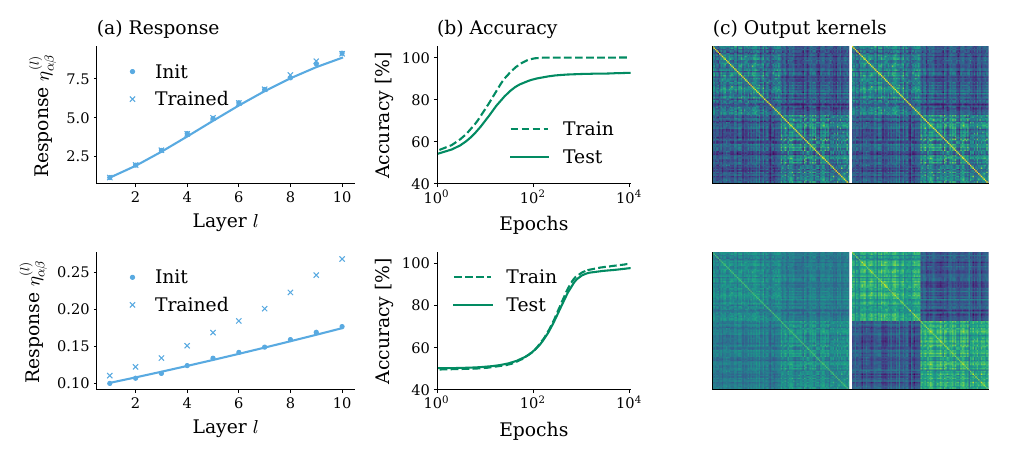}\caption{Behavior of the response function in trained networks. We show results
for $\rho=1.0$ (upper row) and close to optimal scaling $\rho=0.3\approx1/\sqrt{L}$
(lower row). (a) For the residual response $\eta_{\alpha\beta}^{(l)}$
of the covariances, we show empirical results at initialization (dots)
and after training (crosses) as well as theoretical predictions from
the network prior (solid lines). While the response is almost unchanged
after training for $\rho=1.0$, the response is increased after training
for close to optimal residual scaling $\rho=0.3\approx1/\sqrt{L}$.
The empirical values are measured by finite differences of \eqref{eq:response_finite_diff}
with $\epsilon=10^{-6}$ and rescaling the variances to their original
values so that only the covariances are perturbed. (b) The increased
response coincides with slower learning but improved generalization
throughout training time. (c) The empirically measured output kernels
before (left) and after training (right) indicate significantly stronger
network adaptation to the data close to the optimal residual scaling
$\rho=0.3\approx1/\sqrt{L}$. Other parameters: binary classification
on MNIST between digits $0$ and $3$, training data set size $P=100$,
$N_{\text{epochs}}=10^{4}$, $\sigma_{w}^{2}=1.25$, $\sigma_{b}^{2}=0$,
$\ensuremath{d_{\text{in}}=784}$, $d_{\text{out}}=1$, $N=500$,
$L=10$, $\phi=\text{erf}$.\foreignlanguage{american}{\label{fig:trained_networks}}}
\end{figure*}

As shown in the top panels of \prettyref{fig:trained_networks}, we
find that for $\rho=1$ the response is indeed robust with regard
to network changes during training. Comparing the empirically measured
network kernels at initialization (\prettyref{fig:trained_networks}c,
left) and after training (\prettyref{fig:trained_networks}c, right),
we observe that these change only mildly despite the network having
learned to solve the task (\prettyref{fig:trained_networks}b), explaining
the robustness of the response. Choosing optimal residual scaling
$\rho\approx1/\sqrt{L}$ (\prettyref{fig:trained_networks} bottom),
however, leads to a noticeable adaptation of the kernels to the data
after training (\prettyref{fig:trained_networks}c). As a result,
the response across layers increases (\prettyref{fig:trained_networks}a)
and, while network learning is slower across training epochs, the
network generalizes well during the entire training process (\prettyref{fig:trained_networks}b).
Thus, utilizing a residual scaling $\rho=1/\sqrt{L}$ that promotes
better signal propagation across layers results in stronger network
adaptation and generalization properties of the network. These results
are in line with the response functions driving gradient-based learning
\citep{Schoenholz17_01232}, which is also predicted by the theory
of feature learning in feed-forward networks \citep{seroussi23_908,Fischer24_13660}.

We hypothesize that, while the response function increases towards
the optimal residual scaling $\rho=1/\sqrt{L}$, its fundamental dependence
on network hyperparameters such as the depth $L$ does not change
significantly, explaining the success of this scaling choice in trained
networks. Studying this hypothesis analytically requires determining
the Bayesian posterior of the network, which we leave for future work.

\section{Discussion\label{sec:conclusion}}

Understanding signal propagation in neural networks is essential for
a theory of trainability and generalization. Regarding these points,
residual networks have shown to be superior to feed-forward network
\citep{He16_CVPR,He16_630}; scaling the residual branches in ResNets
further amplifies this effect \citep{Szegedy17_4278}. We here derive
a field-theoretic formulation of residual networks that allows us
to determine finite-size corrections in a systematic way. The response
function of residual networks, a measure for the network's sensitivity
to variability in the input and thus network trainability, naturally
appears as the leading-order correction to the NNGP. We show that,
in contrast to feed-forward networks, the response function decays
to zero only asymptotically, consequently allowing information to
propagate to very deep layers, in line with \citet{Yang17_30}. Further,
we show that signal propagation in ResNets is optimal when the signal
distribution utilizes the whole dynamic range of the activation function.
Beyond this range, information is lost due to saturation effects.
Notably, we are able to explain the universality of empirically found
optimal values due to a weak dependence on all network hyperparameters
but the network depth. Thereby, this work sheds light on the interplay
between signal propagation, saturation effects and signal scales in
residual networks. Finally, we show for trained networks that optimal
residual scaling does not only improve signal propagation but also
generalization capabilities.

\paragraph*{Limitations}

Since we study the network prior of residual networks, the obtained
analytical results strictly only apply to networks at initialization.
We provide empirical results for trained networks, however, indicating
that the hyperparameter dependence of the response function may be
robust at training time. Studying the properties of the response after
training analytically requires determining the full Bayesian posterior,
which we leave for future work.

The presented field-theoretic formalism allows the systematic computation
of finite-size corrections from an expansion of the exact action.
We here truncated this expansion at first order, which yields the
response function. In general, higher orders may appear and are computationally
more costly to obtain. By definition, these, however, do not alter
the response function, which was studied in this work.

\paragraph*{Related works}

Understanding the effect of skip connections in residual networks
has been studied using different quantities: Building on the empirical
observation that connections skipping a certain number of fully-connected
layers lead to smaller training errors, \citet{Li17_arxiv} show
that the condition number of the Hessian of the loss function does
not grow with network depth but is depth-invariant. Further, ResNets
achieve improved data separability compared to FFNets as they preserve
angles between samples and thus exhibit less degradation of the ratio
between within-class distance and between-class distance \citep{Furusho19_94}.

While we here explicitly focus on finite-size networks and the implications
of the finite size, other works examine the scaling behavior in the
infinite-size limit. \citet{Yang17_30} explicitly study signal propagation
in residual networks in the infinite-width limit by determining the
decay rate of sample correlations to their fixed point values. They
find a sub-exponential or even polynomial decay, indicating that,
in contrast to feed-forward networks \citep{Poole16_3360,Schoenholz17_01232,Hanin18_neurips},
residual networks are always close to the edge of chaos, leading to
better trainability also at great depth. We recover their result and
go beyond it by investigating the effect of scaling the residual branch
on signal propagation. The closeness to the edge of chaos they describe
can be linked to the long depth scale and increased amplitude of the
response function we demonstrate here.

Neural ordinary differential equations (neural ODEs) \citep{Chen18}
consider the limit of infinite depth for residual networks, yielding
a set of differential equations describing the network dynamics. \citet{Cohen21_2039}
find three different scaling regimes of neural ODEs depending on the
choice of architecture and activation function. \citeauthor{Barboni22_16385}
provide global convergence guarantees for gradient descent and optimal
transport training \citep{Barboni22_16385,Barboni24_arxiv}. \citet{Chen24_iclr}
derive generalization bounds beyond the NTK regime. \citet{Li24_2835}
find a connection between the infinite-depth and -width limits for
residual networks with a scaling by square root of the inverse depth
on the residual branches and feed-forward networks with activation
functions scaled by the same factor, building on the $1/\sqrt{\mathrm{depth}}$
scaling for which we here derive the theoretical basis. \citet{Marion24_arxiv}
find a strong relationship between residual scaling and the regularity
of the weights for obtaining non-trivial dynamics in residual networks,
impacting performance for initialized and trained networks.

The residual scaling parameter affects the behavior of gradients in
ResNets: Smaller scaling values reduce the whitening of gradients
with increasing depth \citep{Balduzzi17_342}. \citet{Zaeemzadeh21_3980}
show that skip connections lead to norm-preservation of the gradients
during backpropagation by shifting the singular values closer to one;
norm-preservation in turn improves trainability and generalization.
Regarding the problem of vanishing or exploding gradients, \citet{Ling19_105212}
require the singular values of the input-output Jacobian to be of
order one, suggesting a scaling of the weight variance by the square
root of the inverse depth. While their result holds only in the double
limit of infinite width and depth, our theoretical predictions apply
to finite-size networks.

Another line of research studies the scaling of the skip connection
instead of the residual branch: According to \citet{Zhang24_arxiv},
scaling the skip connection improves semantic feature learning; suggesting
an inverse scaling with depth while observing that the particular
decay scheme is less relevant than the total decay over all layers.
\citet{Doshi23_neurips} investigate critical points in residual networks,
observing only a weak dependence on other hyperparameters such as
weight and bias variance at initialization similar as the weak dependence
explained by the here developed theory; they empirically compute partial
Jacobians while we study the response function that naturally appears
in the presented field-theoretic framework. Scaling the skip connections
can straightforwardly be included into this framework; we leave this
for future work.

\paragraph*{Outlook}

The field-theoretic formulation in \prettyref{sec:theory_resnets}
allows the systematic computation of finite-size corrections to the
NNGP result. \citet{Fischer24_13660} compute the Bayesian posterior
network kernel that results from fluctuation corrections to the NNGP
kernels in feed-forward networks. In the future, we plan to investigate
the effect of skip connections on the network posterior in this framework.
Further, \citet{Lindner23_arxiv} employ a related framework for studying
the effect of data variability on the network posterior, which is
also an interesting avenue of future research for residual networks.

\begin{acknowledgments}
We thank Peter Bouss and Claudia Merger for helpful discussions. This
work was partly supported by the German Federal Ministry for Education
and Research (BMBF Grant 01IS19077A to Jülich and BMBF Grant 01IS19077B
to Aachen) and funded by the Deutsche Forschungsgemeinschaft (DFG,
German Research Foundation) - 368482240/GRK2416, the Excellence Initiative
of the German federal and state governments (ERS PF-JARA-SDS005),
and the Helmholtz Association Initiative and Networking Fund under
project number SO-092 (Advanced Computing Architectures, ACA). The
authors gratefully acknowledge the computing time granted by the JARA
Vergabegremium and provided on the JARA Partition part of the supercomputer
JURECA at Forschungszentrum Jülich (computation grant JINB33). Open access publication funded by the Deutsche Forschungsgemeinschaft (DFG, German Research Foundation) - 491111487.
\end{acknowledgments}

\begin{appendices}

\begin{widetext}

\section{Network prior for multiple inputs\label{app:prior_multiple_inputs}}

We here calculate an analytical expression for the network prior $p(Y\vert X)$,
derived in \prettyref{sec:theory_resnets} of the main text for a
single input, for multiple inputs $X=(x_{\alpha})_{\alpha=1,\ldots,P}$
and network outputs $Y=(y_{\alpha})_{\alpha=1,\ldots,P}$:
\begin{equation}
p(Y\vert X)=\int\mathrm{d}\theta\,\prod_{\alpha}\,p(y_{\alpha}\vert x_{\alpha},\theta)\,p(\theta).\label{eq:network_prior_resnets-1}
\end{equation}
For fixed network parameters $\theta$, the probability $p(Y\vert X,\theta)$
is given by enforcing the network model with Dirac $\delta$-distributions
as
\begin{align}
p(Y\vert X,\theta) & =\prod_{\alpha}\int\mathrm{d}h_{\alpha}^{(0)}\dots\int\mathrm{d}h_{\alpha}^{(L)}\,\delta(h_{\alpha}^{(0)}-W^{\text{in}}x_{\alpha}-b^{\text{in}})\nonumber \\
 & \qquad\times\prod_{l=1}^{L}\delta(h_{\alpha}^{(l)}-h_{\alpha}^{(l-1)}-\rho W^{(l)}\phi(h_{\alpha}^{(l-1)})-\rho b^{(l)})\label{eq:prior_delta_enforced_resnets-1}\\
 & \qquad\times\delta(y_{\alpha}-W^{\text{out}}\phi(h_{\alpha}^{(L)})-b^{\text{out}}).\nonumber 
\end{align}
In contrast to the case of a single input $x_{\alpha}$ in the main
text, we here need to multiply across all $P$ inputs ${X=(x_{\alpha})_{\alpha=1,\ldots,P}}$.

\subsection*{Marginalization over network parameters}

We write the marginalization over network parameters
\begin{align}
p(Y\vert X) & =\prod_{\alpha}\int\mathrm{d}h_{\alpha}^{(0)}\dots\int\mathrm{d}h_{\alpha}^{(L)}\,\langle\delta(h_{\alpha}^{(0)}-W^{\text{in}}x_{\alpha}-b^{\text{in}})\rangle_{\left\{ W^{\text{in}},\,b^{\text{in}}\right\} }\nonumber \\
 & \qquad\times\prod_{l=1}^{L}\langle\delta(h_{\alpha}^{(l)}-h_{\alpha}^{(l-1)}-\rho W^{(l)}\phi(h_{\alpha}^{(l-1)})-\rho b^{(l)})\rangle_{\left\{ W^{(l)},\,b^{(l)}\right\} }\\
 & \qquad\times\langle\delta(y_{\alpha}-W^{\text{out}}\phi(h_{\alpha}^{(L)})-b^{\text{out}})\rangle_{\left\{ W^{\text{out}},\,b^{\text{out}}\right\} },\nonumber 
\end{align}
where $\left\langle \dots\right\rangle _{\{W,b\}}$ refers to the
Gaussian average over weights $W$ and biases $b$ according to the
priors on the network parameters: For the input layer ${W_{ij}^{\text{in}}\overset{\text{i.i.d.}}{\sim}\mathcal{N}(0,\sigma_{w,\,\text{in}}^{2}/d_{\text{in}}),}\;{b_{i}^{\text{in}}\overset{\text{i.i.d.}}{\sim}\mathcal{N}(0,\sigma_{b,\,\text{in}}^{2}),}\;$for
residual layers $W_{ij}^{(l)}\overset{\text{i.i.d.}}{\sim}\mathcal{N}(0,\sigma_{w}^{2}/N)$,
${b_{i}^{(l)}\overset{\text{i.i.d.}}{\sim}\mathcal{N}(0,\sigma_{b}^{2})},$
and for the readout layer ${W_{ij}^{\text{out}}\overset{\text{i.i.d.}}{\sim}\mathcal{N}(0,\sigma_{w,\,\text{out}}^{2}/N)}$,
${b_{i}^{\text{out}}\overset{\text{i.i.d.}}{\sim}\mathcal{N}(0,\sigma_{b,\,\text{out}}^{2})}$.
We use the Fourier representation of the Dirac $\delta$-distribution
\begin{align}
\delta(z) & =\int\mathrm{d}\tilde{z}\,\exp\big(\tilde{z}^{\T}z\big)
\end{align}
with $\tilde{z}$ the conjugate variable to $z$ and $\tilde{z}^{\T}z=\sum_{i=1}^{N}\tilde{z}_{i}z_{i}$,
where we integrate along the imaginary axis $\int\mathrm{d}\tilde{z}=\prod_{k}\int_{i\mathbb{R}}\frac{\mathrm{d}\tilde{z}_{k}}{2\pi i}$
for notational convenience, and rewrite
\begin{align}
p(Y\vert X) & =\prod_{\alpha}\left\{ \int\mathcal{D}\tilde{y_{\alpha}}\int\mathcal{D}\tilde{h}_{\alpha}\int\mathcal{D}h_{\alpha}\right\} \,\left\langle \exp\Big(\sum_{\alpha,i}\tilde{h}_{i,\alpha}^{(0)}(h_{i,\alpha}^{(0)}-\sum_{j}W_{ij}^{\text{in}}x_{j,\alpha}-b_{i}^{\text{in}})\Big)\right\rangle _{\left\{ W^{\text{in}},\,b^{\text{in}}\right\} }\nonumber \\
 & \qquad\times\prod_{l=1}^{L}\left\langle \exp\Big(\sum_{\alpha,i}\tilde{h}_{i,\alpha}^{(l)}(h_{i,\alpha}^{(l)}-h_{i,\alpha}^{(l-1)}-\rho\sum_{j}W_{ij}^{(l)}\phi_{j,\alpha}^{(l-1)}-\rho b_{i}^{(l)})\Big)\right\rangle _{\left\{ W^{(l)},\,b^{(l)}\right\} }\label{eq:network_prior_delta}\\
 & \qquad\times\left\langle \exp\Big(\sum_{\alpha,i}\tilde{y}_{i,\alpha}(y_{i,\alpha}-\sum_{j}W_{ij}^{\text{out}}\phi_{j,\alpha}^{(L)}-b_{i}^{\text{out}})\Big)\right\rangle _{\left\{ W^{\text{out}},\,b^{\text{out}}\right\} },\nonumber 
\end{align}
where $\int\mathcal{D}h_{\alpha}=\prod_{l=0}^{L}\int\mathrm{d}h_{\alpha}^{(l)}$
and $\int\mathcal{D}\tilde{h}_{\alpha}=\prod_{l=0}^{L}\int\mathrm{d}\tilde{h}_{\alpha}^{(l)}$.
We introduced the shorthand $\phi_{j,\alpha}^{(l-1)}=\phi(h_{j,\alpha}^{(l-1)})$
for brevity. Since the network parameters $\theta_{k}$ are independently
distributed, we can compute the averages for each parameter separately
$\int\mathrm{d}\theta_{k}\,p(\theta_{k})\,\exp\big(z\theta_{k}\big)$,
which yields the moment-generating function of the distribution $p(\theta_{k})$.
For Gaussian distributed parameters $\theta_{k}\sim\mathcal{N}(0,\sigma^{2})$,
this gives 
\begin{align}
\int\mathrm{d}\theta_{k}\,p(\theta_{k})\,\exp\big(z\theta_{k}\big) & =\int\mathrm{d}\theta_{k}\,\frac{1}{\sqrt{2\pi}\sigma}\exp\bigg(-\frac{1}{2\sigma^{2}}\theta_{k}^{2}\bigg)\,\exp\big(z\theta_{k}\big)\\
 & =\int\mathrm{d}\theta_{k}\,\frac{1}{\sqrt{2\pi}\sigma}\exp\bigg(-\frac{1}{2\sigma^{2}}(\theta_{k}-\sigma z)^{2}\bigg)\,\exp\big(\frac{1}{2}\sigma^{2}z^{2}\big)\\
 & =\exp\big(\frac{1}{2}\sigma^{2}z^{2}\big).
\end{align}
For the individual terms in \prettyref{eq:network_prior_delta}, we
then get
\begin{align}
\left\langle \exp\Big(-\sum_{i,j}W_{ij}^{\text{in}}\sum_{\alpha}\tilde{h}_{i,\alpha}^{(0)}x_{j,\alpha}\Big)\right\rangle _{W^{\text{in}}} & =\exp\Big(\frac{1}{2}\frac{\sigma_{w,\,\text{in}}^{2}}{d_{\text{in}}}\,\sum_{i,j}\Big(\sum_{\alpha}\tilde{h}_{i,\alpha}^{(0)}x_{j,\alpha}\Big)^{2}\Big)\\
 & =\exp\Big(\frac{1}{2}\frac{\sigma_{w,\,\text{in}}^{2}}{d_{\text{in}}}\,\sum_{\alpha,\beta}\sum_{i,j}\tilde{h}_{i,\alpha}^{(0)}x_{j,\alpha}\,\tilde{h}_{i,\beta}^{(0)}x_{j,\beta}\Big),\nonumber \\
\left\langle \exp\Big(-\sum_{i}b_{i}^{\text{in}}\sum_{\alpha}\tilde{h}_{i,\alpha}^{(0)})\Big)\right\rangle _{b^{\text{in}}} & =\exp\Big(\frac{1}{2}\sigma_{b,\,\text{in}}^{2}\sum_{i}\Big(\sum_{\alpha}\tilde{h}_{i,\alpha}^{(0)}\Big)^{2}\Big)\\
 & =\exp\Big(\frac{1}{2}\sigma_{b,\,\text{in}}^{2}\sum_{\alpha,\beta}\sum_{i}\tilde{h}_{i,\alpha}^{(0)}\,\tilde{h}_{i,\beta}^{(0)}\Big),\nonumber \\
\left\langle \exp\Big(-\sum_{i,j}W_{ij}^{(l)}\,\rho\sum_{\alpha}\tilde{h}_{i,\alpha}^{(l)}\phi_{j,\alpha}^{(l-1)}\Big)\right\rangle _{W^{(l)}} & =\exp\Big(\frac{1}{2}\frac{\sigma_{w}^{2}}{N}\sum_{i,j}\Big(\rho\sum_{\alpha}\tilde{h}_{i,\alpha}^{(l)}\phi_{j,\alpha}^{(l-1)}\Big)^{2}\Big)\\
 & =\exp\Big(\frac{1}{2}\rho^{2}\frac{\sigma_{w}^{2}}{N}\sum_{\alpha,\beta}\sum_{i,j}\tilde{h}_{i,\alpha}^{(l)}\phi_{j,\alpha}^{(l-1)}\,\tilde{h}_{i,\beta}^{(l)}\phi_{j,\beta}^{(l-1)}\Big),\nonumber 
\end{align}
\begin{align}
\left\langle \exp\Big(-\sum_{i}b_{i}^{(l)}\,\rho\sum_{\alpha}\tilde{h}_{i,\alpha}^{(l)}\Big)\right\rangle _{b^{(l)}} & =\exp\Big(\frac{1}{2}\sigma_{b}^{2}\sum_{i}\Big(\rho\sum_{\alpha}\tilde{h}_{i,\alpha}^{(l)}\Big)^{2}\Big)\nonumber \\
 & =\exp\Big(\frac{1}{2}\rho^{2}\sigma_{b}^{2}\sum_{\alpha,\beta}\sum_{i}\tilde{h}_{i,\alpha}^{(l)}\,\tilde{h}_{i,\beta}^{(l)}\Big),\nonumber \\
\left\langle \exp\Big(-\sum_{i,j}W_{ij}^{\text{out}}\,\sum_{\alpha}\tilde{y}_{i,\alpha}\phi_{j,\alpha}^{(L)}\Big)\right\rangle _{W^{\text{out}}} & =\exp\Big(\frac{1}{2}\frac{\sigma_{w,\,\text{out}}^{2}}{N}\sum_{i,j}\Big(\sum_{\alpha}\tilde{y}_{i,\alpha}\phi_{j,\alpha}^{(L)}\Big)^{2}\Big)\\
 & =\exp\Big(\frac{1}{2}\frac{\sigma_{w,\,\text{out}}^{2}}{N}\sum_{\alpha,\beta}\sum_{i,j}\tilde{y}_{i,\alpha}\phi_{j,\alpha}^{(L)}\,\tilde{y}_{i,\beta}\phi_{j,\beta}^{(L)}\Big),\nonumber \\
\left\langle \exp\Big(-\sum_{i}b_{i}^{\text{out}}\,\sum_{\alpha}\tilde{y}_{i,\alpha}\Big)\right\rangle _{b^{\text{out}}} & =\exp\Big(\frac{1}{2}\sigma_{b,\,\text{out}}^{2}\sum_{i}\,\Big(\sum_{\alpha}\tilde{y}_{i,\alpha}\Big)^{2}\Big)\\
 & =\exp\Big(\frac{1}{2}\sigma_{b,\,\text{out}}^{2}\sum_{\alpha,\beta}\sum_{i}\,\tilde{y}_{i,\alpha}\tilde{y}_{i,\beta}\Big).\nonumber 
\end{align}
To ease notation, we use an implicit summation convention in the following
for lower indices that appear twice as $\sum_{\alpha}\sum_{i}\tilde{h}_{i,\alpha}^{(l)}\,\tilde{h}_{i,\alpha}^{(l)}=\tilde{h}_{i,\alpha}^{(l)}\,\tilde{h}_{i,\alpha}^{(l)}$.
Further, we write $\int\mathcal{D}\tilde{h}=\prod_{\alpha}\int\mathcal{D}\tilde{h}_{\alpha}$.
Rewriting the sums over neuron indices as {\small{}$\sum_{mn}\left[\tilde{h}_{m}\phi_{n}^{(l-1)}\right]^{2}=\tilde{h}^{\T}\tilde{h\,}\left[\phi^{(l-1)}\right]^{\T}\phi^{(l-1)}$},
we overall get for the prior
\begin{align}
p(Y\vert X) & =\int\mathcal{D}\tilde{y}\int\mathcal{D}\tilde{h}\int\mathcal{D}h\,\exp\left(\tilde{y}_{\alpha}^{\T}y_{\alpha}+\frac{1}{2}\frac{\sigma_{w,\,\text{out}}^{2}}{N}\,\tilde{y}_{\alpha}^{\T}\tilde{y}_{\beta}\,\left[\phi_{\alpha}^{(L)}\right]^{\T}\phi_{\beta}^{(L)}+\frac{1}{2}\sigma_{b,\,\text{out}}^{2}\sum_{\alpha,\beta}\tilde{y}_{\alpha}^{\T}\tilde{y}_{\beta}\right)\\
 & \qquad\times\exp\left(\sum_{l=1}^{L}\left[\tilde{h}_{\alpha}^{(l)}\right]^{\T}\left[h_{\alpha}^{(l)}-h_{\alpha}^{(l-1)}\right]\right)\nonumber \\
 & \qquad\times\exp\left(\sum_{l=1}^{L}\left(\frac{1}{2}\rho^{2}\frac{\sigma_{w}^{2}}{N}\,\left[\tilde{h}_{\alpha}^{(l)}\right]^{\T}\tilde{h}_{\beta}^{(l)}\,\left[\phi_{\alpha}^{(l-1)}\right]^{\T}\phi_{\beta}^{(l-1)}+\frac{1}{2}\rho^{2}\sigma_{b}^{2}\,\sum_{\alpha,\beta}\left[\tilde{h}_{\alpha}^{(l)}\right]^{\T}\tilde{h}_{\beta}^{(l)}\right)\right)\nonumber \\
 & \qquad\times\exp\left(\left[\tilde{h}_{\alpha}^{(0)}\right]^{\T}h_{\alpha}^{(0)}+\frac{1}{2}\frac{\sigma_{w,\,\text{in}}^{2}}{d_{\text{in}}}\,\left[\tilde{h}_{\alpha}^{(0)}\right]^{\T}\tilde{h}_{\beta}^{(0)}\,x_{\alpha}{}^{\T}x_{\beta}+\frac{1}{2}\sigma_{b,\,\text{in}}^{2}\,\sum_{\alpha,\beta}\left[\tilde{h}_{\alpha}^{(0)}\right]^{\T}\tilde{h}_{\beta}^{(0)}\right)\nonumber \\
 & \eqqcolon\int\mathcal{D}\tilde{y}\int\mathcal{D}\tilde{h}\int\mathcal{D}h\,\exp\left(\mathcal{S}(Y,\tilde{Y},H,\tilde{H}\vert X)\right).
\end{align}
The action $\mathcal{S}$ consists of three contributions
\begin{align}
\mathcal{S}(Y,\tilde{Y},H,\tilde{H}\vert X) & =\mathcal{S}_{\text{in}}(H^{(0)},\tilde{H}^{(0)}\vert X)+\mathcal{S}_{\text{net}}(H,\tilde{H})+\mathcal{S}_{\text{out}}(Y,\tilde{Y}\vert H^{(L)}),\label{eq:action_total_multiple}
\end{align}
one term from the readin layer containing the dependence on the inputs
$X$ as
\begin{equation}
\mathcal{S}_{\text{in}}(H^{(0)},\tilde{H}^{(0)}\vert X)\coloneqq\left[\tilde{h}_{\alpha}^{(0)}\right]^{\T}h_{\alpha}^{(0)}+\frac{1}{2}\frac{\sigma_{w,\,\text{in}}^{2}}{d_{\text{in}}}\,\left[\tilde{h}_{\alpha}^{(0)}\right]^{\T}\tilde{h}_{\beta}^{(0)}\,x_{\alpha}{}^{\T}x_{\beta}+\frac{1}{2}\sigma_{b,\,\text{in}}^{2}\,\sum_{\alpha,\beta}\left[\tilde{h}_{\alpha}^{(0)}\right]^{\T}\tilde{h}_{\beta}^{(0)},
\end{equation}
one term from the intermediate layers containing the skip connections
\begin{equation}
\mathcal{S}_{\text{net}}(H,\tilde{H})\coloneqq\sum_{l=1}^{L}\left[\tilde{h}_{\alpha}^{(l)}\right]^{\T}\left[h_{\alpha}^{(l)}-h_{\alpha}^{(l-1)}\right]+\frac{1}{2}\rho^{2}\frac{\sigma_{w}^{2}}{N}\,\left[\tilde{h}_{\alpha}^{(l)}\right]^{\T}\tilde{h}_{\beta}^{(l)}\,\left[\phi_{\alpha}^{(l-1)}\right]^{\T}\phi_{\beta}^{(l-1)}+\frac{1}{2}\rho^{2}\sigma_{b}^{2}\,\sum_{\alpha,\beta}\left[\tilde{h}_{\alpha}^{(l)}\right]^{\T}\tilde{h}_{\beta}^{(l)},
\end{equation}
and one term for the readout layer containing the network outputs
$Y$ as
\begin{equation}
\mathcal{S}_{\text{out}}(Y,\tilde{Y}\vert H^{(L)})\coloneqq\tilde{y}_{\alpha}^{\T}y_{\alpha}+\frac{1}{2}\frac{\sigma_{w,\,\text{out}}^{2}}{N}\,\tilde{y}_{\alpha}^{\T}\tilde{y}_{\beta}\,\left[\phi_{\alpha}^{(L)}\right]^{\T}\phi_{\beta}^{(L)}+\frac{1}{2}\sigma_{b,\,\text{out}}^{2}\,\tilde{y}_{\alpha}^{\T}\tilde{y}_{\beta}.
\end{equation}

\subsection*{Auxiliary variables}

In the above integrals, we can only solve Gaussian terms exactly,
which is limited to interaction terms between $h$ and $\tilde{h}$
up to quadratic terms. In the action $\mathcal{S}$ in \eqref{eq:action_total_multiple}
we have terms $\propto\bigg[\tilde{h}^{(l)}\bigg]^{\T}\tilde{h}^{(l)}\,\bigg[\phi^{(l-1)}\bigg]^{\T}\phi^{(l-1)}$
appearing, which are quartic in $h$ and $\tilde{h}$ for linear activations
$\phi(h)=h$ and of higher order for non-linear activations $\phi$.
To decouple these interaction terms into lower-order ones, we introduce
auxiliary variables
\begin{align}
C_{\alpha\beta}^{(l)} & \coloneqq\begin{cases}
\frac{\sigma_{w,\,\text{in}}^{2}}{d_{\text{in}}}\,(XX^{\T})_{\alpha\beta}+\sigma_{b,\,\text{in}}^{2} & l=0,\\
\rho^{2}\frac{\sigma_{w}^{2}}{N}\,\phi_{\alpha}^{(l-1)}\cdot\phi_{\beta}^{(l-1)}+\rho^{2}\sigma_{b}^{2} & 1\leq l\leq L,\\
\frac{\sigma_{w,\,\text{out}}^{2}}{N}\,\phi_{\alpha}^{(L)}\cdot\phi_{\beta}^{(L)}+\sigma_{b,\,\text{out}}^{2} & l=L+1.
\end{cases}
\end{align}
To account for the original higher-order interactions between $\tilde{h}$
and $\phi^{(l-1)}$, we enforce the definition of the auxiliary variables
by Dirac $\delta$-distributions and obtain for the network prior
\begin{align}
p(Y\vert X) & =\int\mathcal{D}\tilde{y}\int\mathcal{D}\tilde{h}\int\mathcal{D}h\,\prod_{\alpha,\beta}\int\mathcal{D}C_{\alpha\beta}\,\exp\left(\tilde{y}_{\alpha}^{\T}y_{\alpha}+\frac{1}{2}C_{\alpha\beta}^{(L+1)}\tilde{y}_{\alpha}^{\T}\tilde{y}_{\beta}\right)\,\delta\bigg(C_{\alpha\beta}^{(L+1)}-\frac{\sigma_{w,\,\text{out}}^{2}}{N}\,\left[\phi_{\alpha}^{(L)}\right]^{\T}\phi_{\beta}^{(L)}-\sigma_{b,\,\text{out}}^{2}\bigg)\nonumber \\
 & \qquad\times\exp\left(\sum_{l=1}^{L}\left[\left[\tilde{h}_{\alpha}^{(l)}\right]^{\T}\left[h_{\alpha}^{(l)}-h_{\alpha}^{(l-1)}\right]+\frac{1}{2}C_{\alpha\beta}^{(l)}\left[\tilde{h}_{\alpha}^{(l)}\right]^{\T}\tilde{h}_{\beta}^{(l)}\right]\right)\,\delta\bigg(C_{\alpha\beta}^{(l)}-\rho^{2}\frac{\sigma_{w}^{2}}{N}\,\left[\phi_{\alpha}^{(l-1)}\right]^{\T}\phi_{\beta}^{(l-1)}-\rho^{2}\sigma_{b}^{2}\bigg)\nonumber \\
 & \qquad\times\exp\left(\left[\tilde{h}_{\alpha}^{(0)}\right]^{\T}h_{\alpha}^{(0)}+\frac{1}{2}C_{\alpha\beta}^{(0)}\left[\tilde{h}_{\alpha}^{(0)}\right]^{\T}\tilde{h}_{\beta}^{(0)}\right)\,\delta\bigg(C_{\alpha\beta}^{(0)}-\frac{\sigma_{w,\,\text{in}}^{2}}{d_{\text{in}}}\,x_{\alpha}\cdot x_{\beta}-\sigma_{b,\,\text{in}}^{2}\bigg),
\end{align}
with $\int\mathcal{D}C_{\alpha\beta}=\prod_{l=0}^{L+1}\int\mathcal{D}C_{\alpha\beta}^{(l)}$.
We rewrite the Dirac $\delta$-distributions using their Fourier representation
\begin{equation}
\delta(C_{\alpha\beta}^{(l)})=\int_{i\R}\frac{\mathrm{d}\tilde{C}_{\alpha\beta}^{(l)}}{2\pi i}\,\exp\big(\tilde{C}_{\alpha\beta}^{(l)}\,C_{\alpha\beta}^{(l)}\big),
\end{equation}
introducing conjugate variables $\tilde{C}_{\alpha\beta}^{(l)}$ to
the auxiliary variables $C_{\alpha\beta}^{(l)}$, as
\begin{align}
\delta\bigg(d_{\text{in}}C_{\alpha\beta}^{(0)}-\sigma_{w,\,\text{in}}^{2}\,x_{\alpha}\cdot x_{\beta}-d_{\text{in}}\sigma_{b,\,\text{in}}^{2}\bigg) & =\int_{i\R}\frac{\mathrm{d}\tilde{C}_{\alpha\beta}^{(0)}}{2\pi i}\,\exp\bigg(d_{\text{in}}\tilde{C}_{\alpha\beta}^{(0)}C_{\alpha\beta}^{(0)}-\sigma_{w,\,\text{in}}^{2}\,\tilde{C}_{\alpha\beta}^{(0)}\,x_{\alpha}\cdot x_{\beta}-d_{\text{in}}\sigma_{b,\,\text{in}}^{2}\,\tilde{C}_{\alpha\beta}^{(0)}\bigg),\\
\delta\bigg(NC_{\alpha\beta}^{(l)}-\rho^{2}\sigma_{w}^{2}\,\left[\phi_{\alpha}^{(l-1)}\right]^{\T}\phi_{\beta}^{(l-1)}-N\rho^{2}\sigma_{b}^{2}\bigg) & =\int_{i\R}\frac{\mathrm{d}\tilde{C}_{\alpha\beta}^{(l)}}{2\pi i}\,\exp\bigg(N\tilde{C}_{\alpha\beta}^{(l)}C_{\alpha\beta}^{(l)}-\rho^{2}\sigma_{w}^{2}\,\tilde{C}_{\alpha\beta}^{(l)}\,\left[\phi_{\alpha}^{(l-1)}\right]^{\T}\phi_{\beta}^{(l-1)}-N\rho^{2}\sigma_{b}^{2}\,\tilde{C}_{\alpha\beta}^{(l)}\bigg),\\
\delta\bigg(NC_{\alpha\beta}^{(L+1)}-\sigma_{w,\,\text{out}}^{2}\,\left[\phi_{\alpha}^{(L)}\right]^{\T}\phi_{\beta}^{(L)}-N\sigma_{b,\,\text{out}}^{2}\bigg) & =\int_{i\R}\frac{\mathrm{d}\tilde{C}_{\alpha\beta}^{(L+1)}}{2\pi i}\,\exp\bigg(N\tilde{C}_{\alpha\beta}^{(L+1)}C_{\alpha\beta}^{(L+1)}-\sigma_{w,\,\text{out}}^{2}\,\tilde{C}_{\alpha\beta}^{(L+1)}\,\left[\phi_{\alpha}^{(L)}\right]^{\T}\phi_{\beta}^{(L)}-N\sigma_{b,\,\text{out}}^{2}\,\tilde{C}_{\alpha\beta}^{(L+1)}\bigg).
\end{align}
Reordering the above terms, we obtain
\begin{align}
p(Y\vert X) & =\int\mathcal{D}\tilde{y}\int\mathcal{D}\tilde{h}\int\mathcal{D}h\int\mathcal{D}C\int\mathcal{D}\tilde{C}\,\exp\left(\tilde{y}_{\alpha}^{\T}y_{\alpha}+\frac{1}{2}C_{\alpha\beta}^{(L+1)}\tilde{y}_{\alpha}^{\T}\tilde{y}_{\beta}\right)\nonumber \\
 & \qquad\times\exp\left(\sum_{l=1}^{L}\left[\left[\tilde{h}_{\alpha}^{(l)}\right]^{\T}\left[h_{\alpha}^{(l)}-h_{\alpha}^{(l-1)}\right]+\frac{1}{2}C_{\alpha\beta}^{(l)}\left[\tilde{h}_{\alpha}^{(l)}\right]^{\T}\tilde{h}_{\beta}^{(l)}\right]\right)\nonumber \\
 & \qquad\times\exp\left(\left[\tilde{h}_{\alpha}^{(0)}\right]^{\T}h_{\alpha}^{(0)}+\frac{1}{2}C_{\alpha\beta}^{(0)}\left[\tilde{h}_{\alpha}^{(0)}\right]^{\T}\tilde{h}_{\beta}^{(0)}\right)\nonumber \\
 & \qquad\times\exp\left(-N\sum_{l=0}^{L+1}\nu_{l}\,C_{\alpha\beta}^{(l)}\,\tilde{C}_{\alpha\beta}^{(l)}+\rho^{2}\sigma_{w}^{2}\,\sum_{l=1}^{L}\tilde{C}_{\alpha\beta}^{(l)}\,\left[\phi_{\alpha}^{(l-1)}\right]^{\T}\phi_{\beta}^{(l-1)}+N\rho^{2}\sigma_{b}^{2}\sum_{l=1}^{L}\sum_{\alpha,\beta}\tilde{C}_{\alpha\beta}^{(l)}\right)\nonumber \\
 & \qquad\times\exp\left(\sigma_{w,\,\text{out}}^{2}\,\tilde{C}_{\alpha\beta}^{(L+1)}\left[\phi_{\alpha}^{(L)}\right]^{\T}\phi_{\beta}^{(L)}+N\sigma_{b,\,\text{out}}^{2}\,\sum_{\alpha,\beta}\tilde{C}_{\alpha\beta}^{(L+1)}\right)\nonumber \\
 & \qquad\times\exp\left(\sigma_{w,\,\text{in}}^{2}\,\tilde{C}_{\alpha\beta}^{(0)}\,(XX^{\T})_{\alpha\beta}+d_{\text{in}}\sigma_{b,\,\text{in}}^{2}\sum_{\alpha,\beta}\tilde{C}_{\alpha\beta}^{(0)}\right),
\end{align}
where we write $\int\mathcal{D}C=\prod_{\alpha,\beta}\left\{ \int\mathcal{D}C_{\alpha\beta}\right\} $
and $\int\mathcal{D}\tilde{C}=\prod_{\alpha,\beta}\prod_{l=0}^{L+1}\int_{i\R}\frac{\mathrm{d}\tilde{C}_{\alpha\beta}^{(l)}}{2\pi i}$
and $\nu_{l}=1+\delta_{0l}\,(d_{\text{in}}/N-1)$. The variables $C_{\alpha\beta}^{(l)}$
and $\tilde{C}_{\alpha\beta}^{(l)}$ only couple to sums of $\tilde{h}_{i,\alpha}^{(l)}$
and $\phi_{i,\alpha}^{(l)}$ over all neuron indices $i$, so all
components of $h_{i,\alpha}^{(l)}$ and $\tilde{h}_{i,\alpha}^{(l)}$
are identically distributed. We can thus rewrite the network prior
in terms of scalar variables, pulling out a factor $N$ in place of
all previous scalar products so that
\begin{align}
p(Y\vert X) & =\int\mathcal{D}\tilde{y}\int\mathcal{D}\tilde{h}\int\mathcal{D}h\int\mathcal{D}C\int\mathcal{D}\tilde{C}\,\exp\left(\tilde{y}_{\alpha}^{\T}y_{\alpha}+\frac{1}{2}C_{\alpha\beta}^{(L+1)}\tilde{y}_{\alpha}^{\T}\tilde{y}_{\beta}\right)\nonumber \\
 & \qquad\times\exp\left(N\sum_{l=1}^{L}\left[\tilde{h}_{\alpha}^{(l)}\left[h_{\alpha}^{(l)}-h_{\alpha}^{(l-1)}\right]+\frac{1}{2}\tilde{h}_{\alpha}^{(l)}C_{\alpha\beta}^{(l)}\tilde{h}_{\beta}^{(l)}\right]\right)\nonumber \\
 & \qquad\times\exp\left(N\sum_{\alpha}\tilde{h}_{\alpha}^{(0)}h_{\alpha}^{(0)}+N\frac{1}{2}\sum_{\alpha,\beta}\tilde{h}_{\alpha}^{(0)}C_{\alpha\beta}^{(0)}\tilde{h}_{\beta}^{(0)}\right)\nonumber \\
 & \qquad\times\exp\left(-N\sum_{l=0}^{L+1}\nu_{l}\,C_{\alpha\beta}^{(l)}\,\tilde{C}_{\alpha\beta}^{(l)}+N\rho^{2}\sigma_{w}^{2}\,\sum_{l=1}^{L}\phi_{\alpha}^{(l-1)}\tilde{C}_{\alpha\beta}^{(l)}\phi_{\beta}^{(l-1)}+N\rho^{2}\sigma_{b}^{2}\sum_{l=1}^{L}\sum_{\alpha,\beta}\tilde{C}_{\alpha\beta}^{(l)}\right)\nonumber \\
 & \qquad\times\exp\left(N\sigma_{w,\,\text{out}}^{2}\,\phi_{\alpha}^{(L)}\tilde{C}_{\alpha\beta}^{(L+1)}\phi_{\beta}^{(L)}+N\sigma_{b,\,\text{out}}^{2}\,\sum_{\alpha,\beta}\tilde{C}_{\alpha\beta}^{(L+1)}\right)\nonumber \\
 & \qquad\times\exp\left(\sigma_{w,\,\text{in}}^{2}\,\tilde{C}_{\alpha\beta}^{(0)}\,(XX^{\T})_{\alpha\beta}+d_{\text{in}}\sigma_{b,\,\text{in}}^{2}\sum_{\alpha,\beta}\tilde{C}_{\alpha\beta}^{(0)}\right).
\end{align}
We again write $h^{(l)}$ and $\tilde{h}^{(l)}$ for their scalar
variables.

We aim to write the network prior $p(Y\vert X)$ in terms of the order
parameters $C_{\alpha\beta}^{(l)}$ and its conjugate variables $\tilde{C}_{\alpha\beta}^{(l)}$.
To this end, we move the integrals over $h^{(l)}$ and $\tilde{h}^{(l)}$
into the exponent, yielding
\begin{align}
p(Y\vert X) & =\int\mathcal{D}\tilde{y}\int\mathcal{D}C\int\mathcal{D}\tilde{C}\,\exp\left(\tilde{y}_{\alpha}^{\T}y_{\alpha}+\frac{1}{2}C_{\alpha\beta}^{(L+1)}\tilde{y}_{\alpha}^{\T}\tilde{y}_{\beta}-N\sum_{l=0}^{L+1}\nu_{l}\,C_{\alpha\beta}^{(l)}\,\tilde{C}_{\alpha\beta}^{(l)}\right)\\
 & \qquad\times\exp\Bigg[N\ln\prod_{l=1}^{L}\int\mathcal{D}\tilde{h}^{(l)}\int\mathcal{D}h^{(l)}\,\exp\left(\tilde{h}_{\alpha}^{(l)}\left[h_{\alpha}^{(l)}-h_{\alpha}^{(l-1)}\right]+\frac{1}{2}\tilde{h}_{\alpha}^{(l)}C_{\alpha\beta}^{(l)}\tilde{h}_{\beta}^{(l)}\right)\nonumber \\
 & \qquad\qquad\qquad\;\times\exp\left(\rho^{2}\sigma_{w}^{2}\,\sum_{l=1}^{L}\phi_{\alpha}^{(l-1)}\tilde{C}_{\alpha\beta}^{(l)}\phi_{\beta}^{(l-1)}+\rho^{2}\sigma_{b}^{2}\sum_{l=1}^{L}\sum_{\alpha,\beta}\tilde{C}_{\alpha\beta}^{(l)}\right)\nonumber \\
 & \qquad\qquad\qquad\;\times\exp\left(\sigma_{w,\,\text{out}}^{2}\,\phi_{\alpha}^{(L)}\tilde{C}_{\alpha\beta}^{(L+1)}\phi_{\beta}^{(L)}+\sigma_{b,\,\text{out}}^{2}\,\sum_{\alpha,\beta}\tilde{C}_{\alpha\beta}^{(L+1)}\right)\nonumber \\
 & \qquad\qquad\qquad\;\times\int\mathcal{D}\tilde{h}^{(0)}\int\mathcal{D}h^{(0)}\exp\left(\tilde{h}_{\alpha}^{(0)}h_{\alpha}^{(0)}+\frac{1}{2}\tilde{h}_{\alpha}^{(0)}C_{\alpha\beta}^{(0)}\tilde{h}_{\beta}^{(0)}\right)\nonumber \\
 & \qquad\qquad\qquad\qquad\quad\times\exp\left(\sigma_{w,\,\text{in}}^{2}\,\tilde{C}_{\alpha\beta}^{(0)}\,(XX^{\T})_{\alpha\beta}+d_{\text{in}}\sigma_{b,\,\text{in}}^{2}\sum_{\alpha,\beta}\tilde{C}_{\alpha\beta}^{(0)}\right)\Bigg]\nonumber \\
 & =\int\mathcal{D}\tilde{y}\left\langle \exp\left(\tilde{y}_{\alpha}^{\T}y_{\alpha}+\frac{1}{2}\tilde{y}_{\alpha}^{\T}C_{\alpha\beta}^{(L+1)}\tilde{y}_{\beta}\right)\right\rangle _{C,\tilde{C}}.
\end{align}
The expectation value appearing in the last line is with respect to
the auxiliary variable $C_{\alpha\beta}^{(l)}$ and its conjugate
variable $\tilde{C}_{\alpha\beta}^{(l)}$, which are distributed according
to the auxiliary action $p(C,\tilde{C})\propto\exp(S(C,\tilde{C}))$
given by
\begin{equation}
\mathcal{S}(C,\tilde{C})\coloneqq-N\sum_{l=0}^{L+1}\nu_{l}\,C_{\alpha\beta}^{(l)}\,\tilde{C}_{\alpha\beta}^{(l)}+N\mathcal{W}(\tilde{C}\vert C),
\end{equation}
with
\begin{align}
\mathcal{W}(\tilde{C}\vert C) & \coloneqq\ln\prod_{l=1}^{L}\int\mathcal{D}\tilde{h}^{(l)}\int\mathcal{D}h^{(l)}\,\exp\left(\tilde{h}_{\alpha}^{(l)}\left[h_{\alpha}^{(l)}-h_{\alpha}^{(l-1)}\right]+\frac{1}{2}\tilde{h}_{\alpha}^{(l)}C_{\alpha\beta}^{(l)}\tilde{h}_{\beta}^{(l)}\right)\\
 & \qquad\times\exp\left(\rho^{2}\sigma_{w}^{2}\,\sum_{l=1}^{L}\phi_{\alpha}^{(l-1)}\tilde{C}_{\alpha\beta}^{(l)}\phi_{\beta}^{(l-1)}+\rho^{2}\sigma_{b}^{2}\sum_{l=1}^{L}\sum_{\alpha,\beta}\tilde{C}_{\alpha\beta}^{(l)}\right)\nonumber \\
 & \qquad\times\exp\left(\sigma_{w,\,\text{out}}^{2}\,\phi_{\alpha}^{(L)}\tilde{C}_{\alpha\beta}^{(L+1)}\phi_{\beta}^{(L)}+\sigma_{b,\,\text{out}}^{2}\,\sum_{\alpha,\beta}\tilde{C}_{\alpha\beta}^{(L+1)}\right)\nonumber \\
 & \qquad\times\int\mathcal{D}\tilde{h}^{(0)}\int\mathcal{D}h^{(0)}\exp\left(N\tilde{h}_{\alpha}^{(0)}h_{\alpha}^{(0)}+N\frac{1}{2}\tilde{h}_{\alpha}^{(0)}C_{\alpha\beta}^{(0)}\tilde{h}_{\beta}^{(0)}+\sigma_{w,\,\text{in}}^{2}\,\tilde{C}_{\alpha\beta}^{(0)}\,(XX^{\T})_{\alpha\beta}+d_{\text{in}}\sigma_{b,\,\text{in}}^{2}\sum_{\alpha,\beta}\tilde{C}_{\alpha\beta}^{(0)}\right).\nonumber 
\end{align}
Note that the conjugate variables $\tilde{C}^{(l)}$ are not proper
random variables, but will be integrated out later to obtain the statistics
of the auxiliary variables $C^{(l)}$.

\section{Saddle point approximation\label{app:saddle_point_approx}}

Since the action $\mathcal{S}$ scales with the network width $N$,
we can perform a saddle point approximation for infinite width $N\rightarrow\infty$.
We compute the saddle points $C_{*},\:\tilde{C}_{*}$ using the conditions
\begin{equation}
\frac{\partial\mathcal{S}}{\partial C_{\alpha\beta}^{(l)}}=0,\;\frac{\partial\mathcal{S}}{\partial\tilde{C}_{\alpha\beta}^{(l)}}=0,
\end{equation}
and obtain
\begin{align}
C_{\alpha\beta,*}^{(l)} & =\begin{cases}
\frac{\sigma_{w,\,\text{in}}^{2}}{d_{\text{in}}}(XX^{\T})_{\alpha\beta}+\sigma_{b,\,\text{in}}^{2} & l=0,\\
\rho^{2}\sigma_{w}^{2}\,\langle\phi_{\alpha}^{(l-1)}\,\phi_{\beta}^{(l-1)}\rangle_{p}+\rho^{2}\sigma_{b}^{2} & 1\leq l\leq L,\\
\sigma_{w,\,\text{out}}^{2}\,\langle\phi_{\alpha}^{(L)}\,\phi_{\beta}^{(L)}\rangle_{p}+\sigma_{b,\,\text{out}}^{2} & l=L+1,
\end{cases}\label{eq:saddle_point_multiple}\\
\tilde{C}_{\alpha\beta,*}^{(l)} & =\begin{cases}
\frac{1}{2}\bigg\langle\tilde{h}_{\alpha}^{(l)}\tilde{h}_{\beta}^{(l)}\bigg\rangle_{p}=0 & 0\leq l\leq L,\\
\frac{1}{2}\bigg\langle\tilde{y}_{\alpha}\tilde{y}_{\beta}\bigg\rangle_{p}=0 & l=L+1,
\end{cases}\label{eq:saddle_point_tilde_C}
\end{align}
with $\tilde{y}=\tilde{h}^{(L+1)}$ for notational brevity in the
following and
\begin{align}
\langle\dots\rangle_{p} & =\prod_{l=1}^{L}\int\mathcal{D}\tilde{h}^{(l)}\int\mathcal{D}h^{(l)}\,\dots\,\exp\left(\tilde{h}_{\alpha}^{(l)}\left[h_{\alpha}^{(l)}-h_{\alpha}^{(l-1)}\right]+\frac{1}{2}\tilde{h}_{\alpha}^{(l)}C_{\alpha\beta,*}^{(l)}\tilde{h}_{\beta}^{(l)}\right)\nonumber \\
 & \qquad\qquad\times\int\mathcal{D}\tilde{h}^{(0)}\int\mathcal{D}h^{(0)}\exp\left(\tilde{h}_{\alpha}^{(0)}h_{\alpha}^{(0)}+\frac{1}{2}\tilde{h}_{\alpha}^{(0)}C_{\alpha\beta,*}^{(0)}\tilde{h}_{\beta}^{(0)}\right).
\end{align}
We show that the second moment of the auxiliary variables $\tilde{h}$
in \eqref{eq:saddle_point_tilde_C} vanishes: We obtain the moments
from the moment-generating function defined as
\begin{equation}
\mathcal{Z}[j,\tilde{j}]\coloneqq\int\mathcal{D}h\int\mathcal{D}\tilde{h}\,\exp\big(\mathcal{S}(h,\tilde{h})+j^{\T}h+\tilde{j}^{\T}\tilde{h}\big)
\end{equation}
by taking derivatives with respect to the source term $\tilde{j}$
\begin{equation}
\langle\tilde{h}_{\alpha_{1}}\dots\tilde{h}_{\alpha_{m}}\rangle=\frac{\partial^{m}\mathcal{Z}[j,\tilde{j}]}{\partial\tilde{j}_{\alpha_{1}}\dots\partial\tilde{j}_{\alpha_{m}}}\rvert_{j,\tilde{j}=0},\label{eq:virtual_moments}
\end{equation}
where $\mathcal{S}(h,\tilde{h})$ is given by \eqref{eq:action_total_multiple}.
Due to the normalization of the probability distribution, it holds
that
\begin{equation}
\mathcal{Z}[0,\tilde{j}]=1\quad\forall\tilde{j}.
\end{equation}
Thus, all derivatives in \eqref{eq:virtual_moments} vanish and consequently
also all moments involving only auxiliary variables $\tilde{h}$.

While the input kernel $C^{(0)}=C_{*}^{(0)}$ in \eqref{eq:saddle_point_multiple}
is fixed by the data $C_{\alpha\beta,\ast}^{(0)}=\sigma_{w,\,\text{in}}^{2}/d_{\text{in}}\,(XX^{\T})_{\alpha\beta}+\sigma_{b,\,\text{in}}^{2}$,
all other residual kernels $C_{*}^{(l)}$ are computed self-consistently
with respect to the expectation value involving $C_{*}^{(l)}$. We
can rewrite this average by defining the residual $f^{(l)}\coloneqq h^{(l)}-h^{(l-1)}$
for $1\leq l\leq L$ and $f^{(0)}\coloneqq h^{(0)}$, yielding
\begin{align}
\langle\dots\rangle_{p} & =\prod_{l=0}^{L}\int\mathcal{D}f^{(l)}\int\mathcal{D}\tilde{h}^{(l)}\,\dots\,\exp\left(\tilde{h}_{\alpha}^{(l)}f_{\alpha}^{(l)}+\frac{1}{2}\tilde{h}_{\alpha}^{(l)}C_{\alpha\beta,*}^{(l)}\tilde{h}_{\beta}^{(l)}\right).\label{eq:def_mf_measure_multidim}
\end{align}
We identify the appearing terms with the moment-generating function
of a Gaussian with zero mean and covariance matrix $C_{\alpha\beta,*}^{(l)}$,
from which follows that the residuals $F^{(l)}=(f_{\alpha}^{(l)})_{\alpha=1,\dots,P}\sim\mathcal{N}(0,C_{*}^{(l)})$
are Gaussian distributed with zero mean and covariance matrix $C_{\alpha\beta,*}^{(l)}$.
Since the expectation values in \eqref{eq:saddle_point_multiple}
depend on $h^{(l)},$ we would like to rewrite $\langle\dots\rangle_{p}$
with respect to $h^{(l)}$. Since the signal $h^{(l)}$ decomposes
into a sum of the residuals as $h^{(l)}=\sum_{k=1}^{l}f^{(l)}$, and
the residuals $f^{(l)}$ are independent across $l$, and means and
covariances for independent Gaussians sum up, the signal $h^{(l)}\sim\mathcal{N}(0,K^{(l)})$
is also Gaussian distributed with zero mean and covariance $K^{(l)}=\sum_{k=0}^{l}C_{*}^{(k)}$.
Using the recursion relation $K^{(l)}=K^{(l-1)}+C_{*}^{(l-1)}$, we
recover the NNGP result \citep{Huang20_33,Tirer22_921,Barzilai23}
for residual networks
\begin{align}
C_{\alpha\beta,*}^{(l)} & =\rho^{2}\sigma_{w}^{2}\,\langle\phi_{\alpha}^{(l-1)}\,\phi_{\beta}^{(l-1)}\rangle_{\mathcal{N}(0,K^{(l-1)})}+\rho^{2}\sigma_{b}^{2},\,\mathrm{for}\,1\leq l\leq L,\label{eq:C_l_multidim}\\
K_{\alpha\beta}^{(l)} & =\begin{cases}
\frac{\sigma_{w,\,\text{in}}^{2}}{d_{\text{in}}}(XX^{\T})_{\alpha\beta}+\sigma_{b,\,\text{in}}^{2} & l=0,\\
K_{\alpha\beta}^{(l-1)}+C_{\alpha\beta,*}^{(l-1)} & 1\leq l\leq L,\\
\sigma_{w,\,\text{out}}^{2}\,\langle\phi_{\alpha}^{(L)}\,\phi_{\beta}^{(L)}\rangle_{\mathcal{N}(0,K^{(L)})}+\sigma_{b,\,\text{out}}^{2} & l=L+1.
\end{cases}\label{eq:nngp_resnets_derived_multidim}
\end{align}

\section{Next-to-leading-order Corrections\label{app:leading_corr_multidim}}

We compute the next-to-leading-order corrections to the saddle points
in the previous section. While the auxiliary variables $C^{(l)}$
concentrate to the saddle point $C_{*}^{(l)}$ for infinite width
$N\rightarrow\infty$, they fluctuate around this value for large
but finite network width $N$. To lowest order, these fluctuations
are Gaussian and can be obtained by computing the Hessian of the action
$\mathcal{S}$ at the saddle point
\begin{align}
p(Y|X) & \simeq\int\mathcal{D}\delta C\,\int\mathcal{D}\delta\tilde{C}\,\exp\Big(\frac{1}{2}(\delta C,\delta\tilde{C})^{\T}\mathcal{S}^{(2)}\,(\delta C,\delta\tilde{C})\Big)\label{eq:fluct_corr}\\
 & =\int\mathcal{D}\delta C\,\int\mathcal{D}\delta\tilde{C}\,\exp\Big(-\frac{1}{2}(\delta C,\delta\tilde{C})^{\T}\mathcal{G}^{-1}\,(\delta C,\delta\tilde{C})\Big),\nonumber 
\end{align}
where we write $\delta C=C-C^{*},\,\delta\tilde{C}=\tilde{C}-\tilde{C}^{*}$
and the negative inverse of the Hessian corresponds to their covariance
\begin{equation}
\mathcal{G}\coloneqq-(\mathcal{S}^{(2)})^{-1}\eqqcolon\left(\begin{array}{cc}
\langle\delta C\,\delta C\rangle & \ensuremath{\langle\delta C\,\delta\tilde{C}\rangle}\\
\ensuremath{\langle\delta\tilde{C}\,\delta C\rangle} & \langle\delta\tilde{C}\,\delta\tilde{C}\rangle
\end{array}\right)
\end{equation}
with $\mathcal{G}$ being the propagator. We evaluate all terms at
the saddle point; thus there is no linear term in \eqref{eq:fluct_corr}
and all expectation values that appear in the following are with respect
to the Gaussian measure $\langle\dots\rangle_{p}$ in \eqref{eq:def_mf_measure_multidim}.

The diagonal entries of the Hessian compute to
\begin{align}
\frac{\partial^{2}}{\partial C_{\alpha\beta}^{(l)}\partial C_{\gamma\beta}^{(k)}}\mathcal{S}\rvert_{(C_{*},\tilde{C}_{*})} & =\frac{1}{4}\bigg\langle\tilde{h}_{\alpha}^{(l)}\tilde{h}_{\beta}^{(l)},\tilde{h}_{\gamma}^{(k)}\tilde{h}_{\delta}^{(k)}\bigg\rangle_{p}^{c}=0,\label{eq:hessian_auxiliary}\\
\frac{\partial^{2}}{\partial\tilde{C}_{\alpha\beta}^{(l)}\partial\tilde{C}_{\gamma\delta}^{(k)}}\mathcal{S}\rvert_{(C_{*},\tilde{C}_{*})} & =\frac{\partial}{\partial\tilde{C}_{\alpha\beta}^{(l)}}\left(N[\delta_{k,L+1}+(1-\delta_{k,L+1})\rho^{2}]\,\sigma_{w}^{2}\langle\phi_{\gamma}^{(k-1)}\phi_{\delta}^{(k-1)}\rangle_{p}+\text{const.}(\tilde{C})\right)\nonumber \\
 & =\delta_{L0}N\left[\sigma_{w,\,\text{in}}^{2}\,(XX^{\T})_{\alpha\beta}+d_{\text{in}}\right]\,[\delta_{k,L+1}+(1-\delta_{k,L+1})\rho^{2}]\,\sigma_{w}^{2}\langle\phi_{\gamma}^{(k-1)}\phi_{\delta}^{(k-1)}\rangle_{p}\\
 & \qquad+N\sigma_{w}^{4}\,1_{l>0}1_{k>0}\,\langle\phi_{\alpha}^{(l-1)}\phi_{\beta}^{(l-1)},\phi_{\gamma}^{(k-1)}\phi_{\delta}^{(k-1)}\rangle_{p}^{c}\nonumber \\
 & \qquad\qquad\times\begin{cases}
\rho^{4} & k,l\neq L+1,\\
\rho^{2} & k\neq l=L+1\vee l\neq k=L+1,\\
1 & \text{else},
\end{cases}\nonumber 
\end{align}
where $1_{l>0}$ denotes the indicator function. We write $\langle\dots\rangle^{c}$
for connected correlations (cumulants) defined as
\begin{align}
\langle z_{\alpha}z_{\beta},z_{\gamma}z_{\delta}\rangle_{p}^{c} & =\langle z_{\alpha}z_{\beta}z_{\gamma}z_{\delta}\rangle_{p}-\langle z_{\alpha}z_{\beta}\rangle_{p}\,\langle z_{\gamma}z_{\delta}\rangle_{p}.
\end{align}
The average over the auxiliary variables $\tilde{h}$ in \eqref{eq:hessian_auxiliary},
where we again use the notation $\tilde{y}=\tilde{h}^{(L+1)}$ for
brevity, vanishes due to the normalization of the probability distribution
as explained in detail in the previous section.

The off-diagonal terms compute to
\begin{align}
 & \frac{\partial^{2}}{\partial C_{\alpha\beta}^{(l)}\partial\tilde{C}_{\gamma\delta}^{(k)}}\mathcal{S}\rvert_{(C_{*},\tilde{C}_{*})}\nonumber \\
 & =-N\nu_{l}\delta_{kl}+N\,1_{k>0}\,\sigma_{w}^{2}\frac{\partial}{\partial C_{\alpha\beta}^{(l)}}\langle\phi_{\gamma}^{(k-1)}\phi_{\delta}^{(k-1)}\rangle_{p}\times\begin{cases}
\rho^{2} & k\leq L\\
1 & k=L+1
\end{cases}\nonumber \\
 & =-N\nu_{l}\delta_{kl}+N\,1_{k>0}\,\sigma_{w}^{2}\frac{\partial}{\partial K_{\alpha\beta}^{(k-1)}}\langle\phi_{\gamma}^{(k-1)}\phi_{\delta}^{(k-1)}\rangle_{\mathcal{N}(0,K^{(k-1)})}\,\frac{\partial}{\partial C_{\alpha\beta}^{(l)}}K_{\alpha\beta}^{(k-1)}\times\begin{cases}
\rho^{2} & k\leq L\\
1 & k=L+1
\end{cases}\nonumber \\
 & =-N\nu_{l}\delta_{kl}+N\,\delta_{(\alpha\beta),(\gamma\delta)}\,1_{k>0}\,\sigma_{w}^{2}\langle\left[\phi_{\alpha}^{(k-1)}\right]^{\prime}\left[\phi_{\beta}^{(k-1)}\right]^{\prime}+\delta_{\alpha\beta}\left[\phi_{\alpha}^{(k-1)}\right]^{\prime\prime}\left[\phi_{\beta}^{(k-1)}\right]\rangle_{\mathcal{N}(0,K^{(k-1)})}\,1_{k>l}\times\begin{cases}
\rho^{2} & k\leq L\\
1 & k=L+1
\end{cases}\label{eq:offdiag_hess-1}
\end{align}
where we used Price's theorem \citep{Price58_69,PapoulisProb4th}
\begin{equation}
\partial_{K_{\alpha\beta}}\langle\phi(h_{\gamma})\phi(h_{\delta})\rangle_{h\sim\mathcal{N}(0,K)}=\langle\partial_{h_{\alpha}}\partial_{h_{\beta}}\phi(h_{\gamma})\phi(h_{\delta})\rangle_{h\sim\mathcal{N}(0,K)}
\end{equation}
from the third to fourth line (see \prettyref{app:prices_theorem}
for details). The condition $k>l$ enforced by the indicator function
$1_{k>l}$ results from the appearing factor $\partial_{C^{(l)}}K^{(k-1)}$
in the derivative, because the network kernel $K^{(k-1)}$ only depends
on the residual kernels $C^{(l)}$ with $l<k$.

To calculate the negative inverse of the Hessian, we write
\begin{align}
\mathcal{S}^{(2)} & =\left(\begin{array}{cc}
\frac{\partial^{2}}{\partial C^{2}}\mathcal{S} & \frac{\partial^{2}}{\partial C\,\partial\tilde{C}}\mathcal{S}\\
\frac{\partial^{2}}{\partial\tilde{C}\,\partial C}\mathcal{S} & \frac{\partial^{2}}{\partial\tilde{C}^{2}}\mathcal{S}
\end{array}\right)\eqqcolon\left(\begin{array}{cc}
\mathcal{S}_{11} & \mathcal{S}_{12}\\
\mathcal{S}_{21} & \mathcal{S}_{22}
\end{array}\right).
\end{align}
Using the block structure and the fact that $\mathcal{S}_{11}=0$,
we get
\begin{align}
\mathcal{G}_{11} & =\mathcal{G}_{12}\,\mathcal{S}_{22}\,\mathcal{G}_{21},\label{eq:delta11-1}\\
\mathcal{G}_{12} & =-\mathcal{S}_{21}^{-1},\\
\mathcal{G}_{22} & =0.
\end{align}

\subsection*{Response function}

The general forward response function is given by
\begin{equation}
\Delta_{12}^{lm,(\alpha\beta),(\gamma\delta)}\coloneqq N\langle C_{(\alpha\beta)}^{(l)}\,\tilde{C}_{(\gamma\delta)}^{(m)}\rangle=N\mathcal{G}_{12}^{(lm),(\alpha\beta),(\gamma\delta)},
\end{equation}
which measures the response of the residual kernel $C_{(\alpha\beta)}^{(l)}$
in layer $l$ to a perturbation in layer $m$ mitigated by the conjugate
variable $\tilde{C}_{(\gamma\delta)}^{(m)}$ . We calculate it as
the negative inverse of the off-diagonal block matrix $\mathcal{S}_{21}$.
Due to the forward-dependence of the kernels $K^{(l)}$, $\mathcal{S}_{21}$
is a lower triangular matrix and we get its inverse from forward propagation
\begin{align}
\mathcal{G}_{12}^{(lm),(\alpha\beta),(\gamma\delta)} & =N^{-1}\nu_{l}^{-1}\delta_{lm}+1_{l>0}\,\delta_{(\alpha\beta),(\gamma\delta)}\,\sigma_{w}^{2}\langle\left[\phi_{\alpha}^{(k-1)}\right]^{\prime}\left[\phi_{\beta}^{(k-1)}\right]^{\prime}+\delta_{\alpha\beta}\left[\phi_{\alpha}^{(k-1)}\right]^{\prime\prime}\left[\phi_{\beta}^{(k-1)}\right]\rangle_{\mathcal{N}(0,K^{(l-1)})}\nonumber \\
 & \qquad\qquad\qquad\qquad\qquad\qquad\qquad\times\sum_{k=0}^{l-1}\mathcal{G}_{12}^{(km),(\alpha\beta),(\gamma\delta)}\times\begin{cases}
\rho^{2} & k\leq L\\
1 & k=L+1
\end{cases}.
\end{align}
 Note that despite the double index $(\alpha\beta),(\gamma\delta)$,
the response is always a function of covariance entries $(\alpha\beta)$
due to the appearing $\delta_{(\alpha\beta),(\gamma\delta)}$ but
depends on all other elements $(\gamma\delta)$ via the appearing
expectation value.

We are ultimately interested in the response function with respect
to the network input as a measure for network trainability. Since
$\Delta_{12}^{(lm),(\alpha\beta),(\gamma\delta)}$ is with respect
to the residual kernels $C^{(l)}$, we define the residual response
for all intermediate network layers $1\leq l\leq L$ as
\begin{equation}
\eta_{\alpha\beta}^{(l)}\coloneqq\Delta_{12}^{(l,0),(\alpha\beta),(\gamma\delta)}
\end{equation}
and obtain
\begin{equation}
\eta_{\alpha\beta}^{(l)}=\delta_{(\alpha\beta),(\gamma\delta)}\,\rho^{2}\sigma_{w}^{2}\langle\left[\phi_{\alpha}^{(l-1)}\right]^{\prime}\left[\phi_{\beta}^{(l-1)}\right]^{\prime}+\delta_{\alpha\beta}\left[\phi_{\alpha}^{(l-1)}\right]^{\prime\prime}\left[\phi_{\beta}^{(l-1)}\right]\rangle_{\mathcal{N}(0,K^{(l-1)})}\,\sum_{k=0}^{l-1}\eta_{\alpha\beta}^{(k)}
\end{equation}
with initial condition $\eta_{\alpha\beta}^{(0)}=N/d_{\mathrm{in}}$.
Due to their additive structure the response function of the kernels
$K^{(l)}$ is given by 
\begin{equation}
\chi_{\alpha\beta}^{(l)}\coloneqq\sum_{k=0}^{l}\eta_{\alpha\beta}^{(k)}.
\end{equation}
Finally, the output response $\chi_{\alpha\beta}^{\text{out}}$ is
then given by
\begin{equation}
\chi_{\alpha\beta}^{\text{out}}=\sigma_{w,\,\text{out}}^{2}\,\langle\left[\phi_{\alpha}^{(L)}\right]^{\prime}\left[\phi_{\beta}^{(L)}\right]^{\prime}+\delta_{\alpha\beta}\left[\phi_{\alpha}^{(L)}\right]^{\prime\prime}\left[\phi_{\beta}^{(L)}\right]\rangle_{\mathcal{N}(0,K^{(L)})}\,\sum_{k=0}^{L}\,\eta_{\alpha\beta}^{(k)}.
\end{equation}

\subsection*{Kernel fluctuations}

Using \eqref{eq:delta11-1}, we get for the diagonal term
\begin{equation}
\mathcal{G}_{11}^{(lm)}=\sum_{k,n}\mathcal{G}_{12}^{(lk)}\,\mathcal{S}_{22}^{(kn)}\,\mathcal{G}_{21}^{(nm)}.
\end{equation}
This quantity describes the Gaussian fluctuations of the residual
kernels $C^{(l)}$ in networks of finite width $N$ around the NNGP
value: 
\begin{equation}
C^{(l)}=C_{*}^{(l)}+\delta C^{(l)}\text{ with }\delta C^{(l)}\sim\mathcal{N}(0,\mathcal{G}_{11}^{(ll)}).
\end{equation}
 The fluctuations for the network kernels $K^{(l)}$ can be once again
computed based on their additive nature as 
\begin{equation}
K^{(l)}=\sum_{k=0}^{l}C^{(k)}=\sum_{k=0}^{l}C_{*}^{(k)}+\sum_{k=0}^{l}\delta C^{(k)},
\end{equation}
yielding
\[
\delta K^{(l)}\sim\mathcal{N}(0,\sum_{k=0}^{l}\mathcal{G}_{11}^{(kk)}).
\]
Kernel fluctuations can be used to compute finite-width corrections
to quantities such as the posterior kernels, generalization error,
etc \citep{Lindner23_arxiv,Fischer24_13660,Rubin25_accepted}.

\section{Price's theorem\label{app:prices_theorem}}

Consider an expectation value of $f:\mathbb{R}^{N}\rightarrow\mathbb{R}$
over centered jointly Gaussian distributed $x_{i}$ with covariance
$C$
\begin{align}
\langle f(x)\rangle_{x\sim\mathcal{N}(0,C)}.
\end{align}
We assume that $f(x)$ grows slower than $e^{x_{i}^{2}}$ for large
$x_{i}$. Rewriting the Gaussian $\mathcal{N}(0,C)$ in terms of its
Fourier transform $\mathcal{N}(0,C)=\Big\{\prod_{j}\int_{-i\infty}^{i\infty}\,\frac{d\tilde{x}_{j}}{2\pi i}\Big\}\exp\big(-x^{\T}\tilde{x}+\frac{1}{2}\tilde{x}^{\T}C\tilde{x}\big)$
one obtains
\begin{align}
\langle f(x)\rangle_{x\sim\mathcal{N}(0,C)} & =\prod_{j}\Big\{\int_{-\infty}^{\infty}dx_{j}\,\int_{-i\infty}^{i\infty}\,\frac{d\tilde{x}_{j}}{2\pi i}\Big\}\,f(x)\,\exp\big(-x^{\T}\tilde{x}+\frac{1}{2}\tilde{x}^{\T}C\tilde{x}\big),
\end{align}
which yields the property
\begin{align}
\frac{\partial}{\partial C_{kl}}\langle f(x)\rangle_{x\sim\mathcal{N}(0,C)} & =\prod_{j}\Big\{\int_{-\infty}^{\infty}dx_{j}\,\int_{-i\infty}^{i\infty}\,\frac{d\tilde{x}_{j}}{2\pi i}\Big\}\,f(x)\,\frac{1}{2}\,\tilde{x}_{k}\tilde{x}_{l}\,\exp\big(-x^{\T}\tilde{x}+\frac{1}{2}\tilde{x}^{\T}C\tilde{x}\big).
\end{align}
One notices that one may replace both occurrences of $\tilde{x}_{i}\to-\partial/\partial x_{i}$
under the integral: integrating by parts twice and using the assumption
that $f$ grows slower than $e^{x_{i}^{2}}$ for large $x_{i}$ so
that boundary terms vanish, one obtains
\begin{align}
\frac{\partial}{\partial C_{kl}}\langle f(x)\rangle_{x\sim\mathcal{N}(0,C)} & =\prod_{j}\Big\{\int_{-\infty}^{\infty}dx_{j}\,\int_{-i\infty}^{i\infty}\,\frac{d\tilde{x}_{j}}{2\pi i}\Big\}\,\frac{1}{2}\,\Big\{\frac{\partial}{\partial x_{k}}\,\frac{\partial}{\partial x_{l}}\,f(x)\Big\}\,\exp\big(x^{\T}\tilde{x}+\frac{1}{2}\tilde{x}^{\T}C\tilde{x}\big)\nonumber \\
 & =\frac{1}{2}\,\big\langle f_{kl}^{(2)}\big\rangle_{x\sim\mathcal{N}(0,C)},\label{eq:price_general-2}
\end{align}
where $f_{kl}^{(2)}$ is the Hessian of $f$. This expression is known
as Price's theorem \citep{Price58_69,PapoulisProb4th}. Note that
sometimes the theorem is only stated for derivatives by $C_{k\neq l}$
only.

\section{Maximum entropy condition for optimal scaling\label{app:bukva_max_ent}}

We here derive an alternative condition for optimal signal variance,
building on \citet{Bukva23_arxiv} who proposed this method to study
trainability in feed-forward networks. Their conjecture is that networks
with signal distributions that are approximately uniform, or put differently
maximally entropic, are more expressive.

For wide networks, the signal distribution of internal layers is approximately
Gaussian
\begin{equation}
p(h;\sigma^{2})=\frac{1}{\sqrt{2\pi\sigma^{2}}}\exp\left(-\frac{1}{2\sigma^{2}}h^{2}\right),
\end{equation}
considering only a scalar component $h$ here as all components $h_{i}$
are independently and identically distributed.

We here focus on the readout layer. The distribution of the post-activation
$y=\phi(h)$ is then
\begin{equation}
p(y;\sigma^{2})=\frac{1}{\sqrt{2\pi\sigma^{2}}\phi^{\prime}(\phi^{-1}(x))}\exp\left(-\frac{1}{2\sigma^{2}}\phi^{-1}(zy)\right).
\end{equation}
For $\phi=\text{erf}$, the post-activation is approximately bounded
by $y\in[-1,1]$. Thus, we compute the Kullback-Leibler divergence
between the distribution of the post-activation and a uniform distribution
on that interval
\begin{align}
D_{\text{KL}}(p_{\text{uni}}\vert p_{\phi}) & =\int_{-1}^{1}\mathrm{d}y\,p_{\text{uni}}(y)\,[\ln p_{\text{uni}}(y)-\ln p_{\phi}(y)]\\
 & =\int_{-1}^{1}\mathrm{d}y\,\frac{1}{2}\ln\left(\frac{1}{2}\right)+\frac{1}{2}\frac{1}{2\sigma^{2}}\phi^{-1}(y)+\frac{1}{2}\ln\left(\sqrt{2\pi}\sigma\phi^{\prime}(\phi^{-1}(y))\right)\nonumber \\
 & =\ln\left(\frac{1}{2}\right)+\frac{1}{2}\int_{-1}^{1}\mathrm{d}y\,\frac{1}{2\sigma^{2}}\phi^{-1}(y)^{2}+\ln\left(\sqrt{2\pi}\sigma\frac{2}{\sqrt{\pi}}\exp(-\phi^{-1}(y)^{2})\right)\nonumber \\
 & =\ln\left(\frac{1}{2}\right)+\ln(\sqrt{8}\sigma)+\frac{1}{2}\int_{-1}^{1}\mathrm{d}y\,\left(\frac{1}{2\sigma^{2}}-1\right)\phi^{-1}(y)^{2}\nonumber \\
 & =\ln\left(\frac{\sqrt{8}}{2}\right)+\ln(\sigma)+\frac{1}{2}\left(\frac{1}{2\sigma^{2}}-1\right)\int_{-\infty}^{\infty}\mathrm{d}h\,\phi^{-1}(\phi(h))^{2}\,\phi^{\prime}(h)\nonumber \\
 & =\ln\left(\sqrt{2}\right)+\frac{1}{2}\ln(\sigma^{2})+\frac{1}{2}\left(\frac{1}{2\sigma^{2}}-1\right)\int_{-\infty}^{\infty}\mathrm{d}h\,h^{2}\,\frac{2}{\sqrt{\pi}}\exp(-h^{2})\nonumber \\
 & =\ln\left(\sqrt{2}\right)+\frac{1}{2}\ln(\sigma^{2})+\frac{1}{2}\left(\frac{1}{2\sigma^{2}}-1\right).
\end{align}
Maximizing the Kullback-Leibler divergence amounts to
\begin{align}
0=\frac{\partial}{\partial\sigma^{2}}D_{\text{KL}}(p_{\text{uni}}\vert p_{\phi}) & =\frac{1}{\sigma^{2}}-\frac{1}{4}\frac{1}{\sigma^{4}},
\end{align}
yielding as the condition for the signal variance before the readout
layer $\sigma^{2}=1/4$. This condition is equivalent to the one in
the \prettyref{sec:signal_prop_scaling} under the assumption that
the dynamic range of the error function is given by $\mathcal{V}=1$.

\section{Additional plots\label{app:add_plots}}

\subsection{Decay of response function\label{app:decay_response}}

Matching the observations by \citet{Yang17_30}, the response function
in ResNets decays sub-exponentially as shown in \prettyref{fig:decay_response}.

\begin{figure}[H]
\centering{}\includegraphics{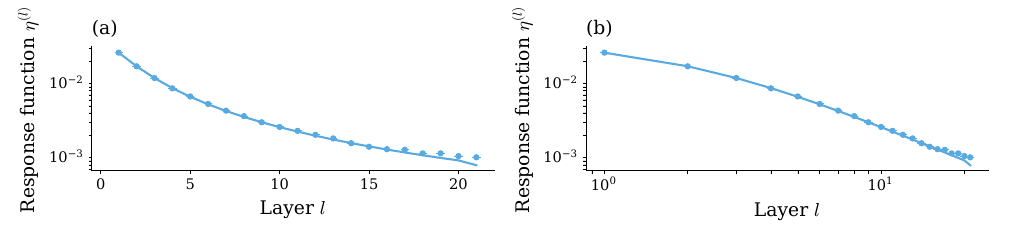}\caption{\foreignlanguage{american}{Log-plot (a) and log-log-plot (b) of the response function $\eta^{(l)}$
for a residual network of depth $L=20$. Dots represent simulations
over $10^{2}$ input samples and $10^{3}$ network initializations,
solid curves show theory values. (a) The decay of the response function
is sub-exponential. (b) In later layers, the decay follows a power
law. Other parameters: $\sigma_{w,\,\text{in}}^{2}=\sigma_{w}^{2}=\sigma_{w,\,\text{out}}^{2}=1.2,\,\sigma_{b,\,\text{in}}^{2}=\sigma_{b}^{2}=\sigma_{b,\,\text{out}}^{2}=0.2,\,d_{\text{in}}=d_{\text{out}}=100,\,N=500,\,\rho=1$,
$\phi=\text{erf}$.\label{fig:decay_response}}}
\end{figure}

\subsection{Input kernels for different tasks}

In \prettyref{sec:signal_prop_scaling}, we study the signal propagation
and scaling behavior in residual networks for different tasks. In
\prettyref{fig:input_kernels}, we show the normalized overlap kernels
$\frac{1}{\max_{\alpha\beta}x_{\alpha}\cdot x_{\beta}}X^{\T}X$ with
$P$ data samples for the studied tasks. For MNIST, we consider binary
classification between $0$ and $3$ with equal number of samples
from both classes $P_{0}=P_{3}=\frac{1}{2}P$. For CIFAR-10, we investigate
binary classification between airplanes and dogs with equal number
of samples from both classes $P_{\text{airplane}}=P_{\text{dog}}=\frac{1}{2}P$.

\begin{figure}[H]
\centering{}\includegraphics{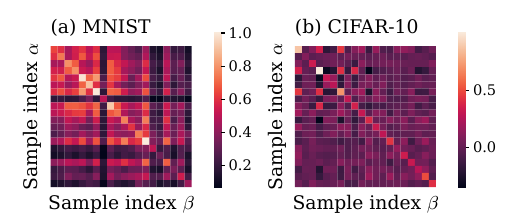}\caption{\foreignlanguage{american}{Normalized overlap kernels of different tasks for $P=20$ samples.
Details on the tasks can be found in the text.\label{fig:input_kernels}}}
\end{figure}

\end{widetext}\end{appendices}

\bibliographystyle{apsrev4-2_prx}
\bibliography{brain,add_to_brain}
\selectlanguage{american}%

\end{document}